\documentclass[superscriptaddress,nobibnotes,amsmath,amssymb,notitlepage,twocolumn,pre,longbibliography ]{revtex4-1}

\usepackage[usenames,dvipsnames]{xcolor}

\usepackage{float}
\usepackage{soul}
\usepackage{amsmath}
\usepackage{amsfonts}
\usepackage{amssymb}
\usepackage{xr-hyper}
\usepackage[colorlinks=true,citecolor=cyan,linkcolor=magenta,filecolor=magenta]{hyperref}
\usepackage[capitalize]{cleveref}
\usepackage{graphicx}
\usepackage{bbold}					
\usepackage[makeroom]{cancel}		
\usepackage{multirow}				
\usepackage[normalem]{ulem}         
\usepackage{array}
\usepackage{mathrsfs}
\usepackage{mathtools}
\usepackage[version=4]{mhchem}      
\usepackage{IEEEtrantools}          
\usepackage{dcolumn}                
\usepackage{physics}                
\usepackage{dsfont}                 
\usepackage{enumerate}              
\usepackage[shortlabels]{enumitem}  
\usepackage{verbatim}

\usepackage{mathdots}

\usepackage{tikz}
\usepackage{subfigure}

\usepackage{graphicx,color,hyperref}
\hypersetup{colorlinks=true, linkcolor=blue, citecolor=blue, urlcolor=blue} 

\graphicspath{{./}}

\newcommand{\beq}[1]{\begin{equation}\label{#1}}
\newcommand{\eep}{\;.\end{equation}}
\newcommand{\eec}{\;,\end{equation}}
\newcommand{\eeq}{\end{equation}}

\newcommand{\om}{\omega}

\DeclareMathAlphabet{\mathcal}{OMS}{cmsy}{m}{n} 

\renewcommand{\vec}[1]{{\bf #1}}

\newcommand{\kv}{\vec{k}}

\newcommand{\nv}{\vec{n}}

\begin{document}

\preprint{APS/123-QED}

\title{Three-dimensional $\mathcal{P}\mathcal{T}$-symmetric topological phases with a Pontryagin index}

\author{Zory Davoyan}

\email{zorydavoyan@gmail.com}

\affiliation{TCM Group, Cavendish Laboratory, Department of Physics, J J Thomson Avenue, Cambridge CB3 0HE, United Kingdom}

\author{Wojciech J. Jankowski}

\affiliation{TCM Group, Cavendish Laboratory, Department of Physics, J J Thomson Avenue, Cambridge CB3 0HE, United Kingdom}
 
\author{Adrien Bouhon}
\affiliation{TCM Group, Cavendish Laboratory, Department of Physics, J J Thomson Avenue, Cambridge CB3 0HE, United Kingdom}
\affiliation{Nordita, Stockholm University and KTH Royal Institute of Technology, Hannes Alfv{\'e}ns v{\"a}g 12, SE-106 91 Stockholm, Sweden}
 
\author{Robert-Jan Slager}
\email{rjs269@cam.ac.uk}
\affiliation{TCM Group, Cavendish Laboratory, Department of Physics, J J Thomson Avenue, Cambridge CB3 0HE, United Kingdom}

\date{\today}

\begin{abstract}
We report on a certain class of three-dimensional topological insulators and semimetals protected by spinless $\mathcal{P}\mathcal{T}$ symmetry, hosting an integer-valued bulk invariant. We show using homotopy arguments that these phases host multigap topology, providing a realization of a single $\mathbb{Z}$ invariant in three spatial dimensions that is distinct from the Hopf index. We identify this invariant with the Pontryagin index, which describes Belavin-Polyakov-Schwartz-Tyupkin (BPST) instantons in particle physics contexts and corresponds to a three-sphere winding number. We study naturally arising multigap linked nodal rings, topologically characterized by split-biquaternion charges, which can be removed by non-Abelian braiding of nodal rings, even without closing a gap. We additionally recast the describing winding number in terms of gauge-invariant combinations of non-Abelian Berry connection elements, indicating relations to Pontryagin characteristic class in four dimensions. These topological configurations are furthermore related to fully non-degenerate multigap phases that are characterized by a pair of winding numbers relating to two isoclinic rotations in the case of four bands and can be generalized to an arbitrary number of bands. From a physical perspective, we also analyze the edge states corresponding to this Pontryagin index as well as their dissolution subject to the gap-closing disorder. Finally, we elaborate on the realization of these novel non-Abelian phases, their edge states and linked nodal structures in acoustic metamaterials and trapped-ion experiments.
\end{abstract}

\maketitle

\section{Introduction}
The study of topological insulators and semimetals provides for an active area of current research that connects various theoretical as well as experimental impetuses~\cite{rmp1,rmp2,Weyl}, offering, amongst others, a condensed matter realization of the $\theta$-vacuum  and according magnetoelectric polarizability, quantum field-theoretic anomalies and axion electrodynamics~\cite{Qi_2008, Axion3,Axion1,Axion4,Axion5,Axion6}. While the inclusion of spatial symmetries, defects and even out-of-equilibrium contexts has provided for an extensive landscape of characterizations~\cite{Clas1,Clas2,Shiozaki14,PointGroupsTI,SchnyderClass, UnifiedBBc, Codefects1, Wi3, RoyZ2,song2018quantitative, Mineev1998topstable}, the past years a rather general viewpoint has emerged. Namely, using readily implementable relations between band representations at high symmetry momenta, general constraint equations that match equivariant K-theory computations~\cite{Clas3} can be derived. 
The emerging classes in momentum space can subsequently be compared to real space band representations~\cite{Zak_EBR1,Zak_EBR2, Zak_EBR3} to discern whether they are compatible with an atomic limit and, accordingly, their topological nature ~\cite{Clas4,Clas5}. Although these symmetry-indicated techniques map out a large fraction of topological insulators and semimetals, the past few years have seen the rise of novel types of topologies that depend on multigap conditions that \textit{a~priori} cannot be captured by these schemes.

These multigap topologies are characterized by finer topological structures and homotopy invariants~\cite{bouhonGeometric2020} that pertain to band spaces (groups of isolated bands)  that in turn depend on multigap conditions. A particular example in this regard is Euler class, being the analog of the Chern number, that arises in systems enjoying $C_2\cal{T}$, that is, two-fold rotations combined with time-reversal symmetry (TRS), or $\cal{PT}$ symmetry, involving parity and TRS. In such scenarios 
band degeneracies residing between different bands carry non-Abelian charges~\cite{Wu1273,BJY_nielsen,bouhon2019nonabelian,bouhonGeometric2020,Tiwari}, being the band structure incarnation of a $\pi$-disclination in a bi-axial nematic~\cite{Kamienrmp,volovik2018investigation, Beekman20171},  and braiding them around in momentum space can result in two-band subspaces that have band nodes with similar, rather than oppositely-valued charges, whose obstruction to be annihilated is directly proportional to the Euler class characterizing that two-band subspace~\cite{bouhon2019nonabelian}. While these new insights have in first stages furnished deeper understanding of finer topologies and the relation to flag manifolds~\cite{bouhonGeometric2020, Ahn2019,Ahn2019SW}, recent advances promise progress in novel quantum geometric structures~\cite{bouhon2023quantum}. More importantly, these multigap topologies are increasingly being related to real physical settings. For example, novel multigap out-of-equilibrium  phases~\cite{slager2022floquet} and in particular quench effects~\cite{Eulerdrive} have been seen in trapped-ion insulators~\cite{Zhao_2022}, while non-Abelian braiding and multigap physics for both bulk and boundary properties have been predicted in phonon~\cite{Peng2021, peng2022multi} as well as electronic spectra of real materials subjected to stress/strain or temperature-induced structural phase transitions~\cite{konye2021,bouhon2019nonabelian,chen2021manipulation}. Finally, these new multigap~topologies are particularly appealing in the context of metamaterials in which an increasing number of theoretical as well as experimental results are being reported~\cite{guo2021experimental, Jiang_meron, Jiang1DExp, bouhon2023second,Jiang_2021,park2022nodal}.

We note that these pursuits fit in a wider research activity that concerns the exploration of phases beyond the ten-fold way~\cite{Kitaev} such as fragile phases~\cite{Ft1} and Hopf insulators~\cite{Hopf_1}, where the first refers to phases in which the topology can be undone by closing the gap with trivial bands (as opposed to a K-theoretical invariant that necessitates a gap closing with a band of opposite charge), while the second type of invariant arises by virtue of the target space under the Hamiltonian mapping. That is, the mapping from a three-dimensional Brillouin zone torus to a two-sphere ($S^2$) target space allows for an identification with a Hopf invariant. We reiterate, however, that multigap phases in principle do not have to be symmetry indicated (all bands can be in the same irreducible representation) and that, while a band gap closing with a trivial band can undo the topology, an ``unbraiding'' process is needed to accomplish this, signaling a different stability~\cite{jankowski2023disorderinduced,wang2023anderson} and characterization. Similarly, with regard to the Hopf map, we stress that, although out-of-equilibrium effects of two-dimensional Euler insulators~\cite{Eulerdrive} or three-dimensional $\cal{PT}$-symmetric insulators (with a four-band target space) can be associated with Hopf maps~\cite{lim2023real}, these maps are generalizations of the standard Hopf map and depend on multigap conditions, that is the partitioning of the bands. 

It is in this setting of refined band partitions and homotopy invariants where we find the subsequent results. In particular, we show that for simple four-band systems the classification of three-dimensional real topological phases can be extended with another type of $\mathbb{Z}$ invariant, which relates to a generalized Pontryagin index representing Belavin-Polyakov-Schwartz-Tyupkin (BPST) instantons in non-Abelian gauge field theories such as $SU(2)$ Yang-Mills theory \cite{Zee2010-ZEEQFT, fradkin}. We demonstrate that this invariant in some sense relates to the $\mathbb{Z}\oplus\mathbb{Z}$-valued Hopf indices characterizing such four-band systems~\cite{lim2023real}, but in fact is a different entity beyond this classification.
Concretely, the index corresponds to the elements of the third homotopy group $\pi_3(S^3) \cong \mathbb{Z}$, describing higher-dimensional winding numbers on the three-sphere $S^3$. Upon adding a real positive tuning parameter $t$ providing an additional dimension in the parameter space, we then establish a link to the so-called characteristic Pontryagin class, a real relative of the second Chern number, accessible in four dimensions. Very interestingly, we also show that this $\mathbb{Z}$ invariant actually characterizes the topology of fully gapped phases (i.e. with complete flag classifying spaces) of even arbitrarily many isolated bands. We moreover provide systematically generated minimal models exhibiting this type of topology, which offers a very direct route towards experimental simulations in optical lattices and metamaterials. Using these models, we also show an interplay between non-Abelian Wilson loops and non-trivial Zak phases~\cite{zak1,Zak2,bouhon2019wilson, Alex_BerryPhase} and edge modes induced at the surfaces and numerically study their robustness to uniform disorder up to the closing of bulk gap. 
As a side result, we find that this type of topology enables braiding of non-Abelian nodal rings in three dimensions, which yields exotic nodal structures in explicit and surprisingly simple Hamiltonians. Accordingly, we propose concrete metamaterial realizations to capture these linking structures as well as the bulk invariant and its bulk-boundary correspondence, thereby impacting active experimental pursuits. 

This paper is organized as follows. In Sec.~\ref{sec::II}, we introduce mathematical definitions and constructs, including classifying spaces, which capture the topology encoded in models with non-trivial Pontryagin index. Section~\ref{sec::III} then presents model realizations of these types of topology, including a minimal one that naturally realizes a non-Abelian linked nodal ring structure. In Sec.~\ref{sec::IV}, we elaborate on the manifestations of the introduced topology, demonstrating bulk-boundary correspondence, associated topological phase transitions, and possible unbraiding mechanisms with and without closing the bulk gap that admit removal of these linked structures, which naturally emerge due to the one-dimensional topology of the flag manifold underlying the algebra of nodal rings. Subsequently, we discuss in Sec. \ref{sec::V} the full flag limit and a multigap invariant on removing the nodal structures, accessing a nontrivial topological phase with a fully non-degenerate band structure, while in Sec.~\ref{sec::VI} we analyze the robustness to disorder of the edge modes induced by the introduced bulk invariants. Finally, we comment on connections with experimental realizations  in Sec.~\ref{sec::VII}, before concluding in Sec.~\ref{sec::VIII}.

\section{Non-Abelian Pontryagin topology}\label{sec::II}
We begin with a general introduction to the non-Abelian real topology realized in the three-dimensional (3D) four-band models proposed in the subsequent section. After introducing the relevant classifying spaces, we then elaborate on the Pontryagin index and its relation to the Pontryagin class related to the realized bulk topology.
 
\subsection{Relevant classifying spaces}
As alluded to above, the topology of the system is fully set by the target space, as induced by the Hamiltonian mapping. This is quantified by the notion of the associated classifying space.
Here, we introduce the classifying spaces relevant for the Pontryagin topology. The classifying space $\mathcal{G}$ is defined to minimally capture the topology of a particular Hamiltonian, which can be induced by the following mappings: $T^d \rightarrow S^d \rightarrow \mathcal{G}$, where the Brillouin zone (BZ) is identified with a $d$-torus, $\text{BZ} \cong T^d$~\cite{SchnyderClass,bouhonGeometric2020, bouhon2023quantum}. In the subsequent we assume the first mapping to be trivial, thereby neglecting possible inducing weak invariants, while the second map is classified by a homotopy group $\pi_d(\mathcal{G})$ capturing all possible nontrivial winding, or topology, of the Hamiltonian map. We require four Bloch bands, spanning a four-dimensional real vector space as a fiber at each crystal momentum $\kv$ in three spatial dimensions, which relates to the topology of a rank-four vector bundle over the BZ hypertorus, $\text{BZ} \cong T^3$, as the base space. We map the BZ to a three-sphere $T^3 \rightarrow S^3$, on which the non-trivial winding of the Hamiltonian captured by the Pontryagin index will be induced by the winding of the isolated band corresponding to the normal bundle of the three-sphere, $NS^3$, with the other three potentially degenerate bands spanning the tangent bundle $TS^3$. The classifying space, and hence topology, is then set by the partitioning of flattened bands.  In particular, on partitioning the system into three-band occupied and single-band unoccupied subspaces,  and assuming a real-valued Hamiltonian due to the presence of spinless spatiotemporal inversion symmetry ($\mathcal{P}\mathcal{T}$~symmetry), the classifying space becomes~\cite{bouhonGeometric2020}
\\
\beq{}
    \mathsf{Gr}_{1,4}(\mathbb{R}) = O(4)/[O(1) \times O(3)] \cong S^3 / \mathbb{Z}_2 \cong \mathbb{RP}^3.
\eeq
\\
Here, we manifestly divide by the group of gauge transformations corresponding to the specific partitioning in the band subspaces. The above manifold corresponds to a real Grassmannian  $\mathsf{Gr}_{k,N}(\mathbb{R})$, where
\\
\beq{}
    \mathsf{Gr}_{k,N}(\mathbb{R}) = O(N)/[O(k) \times O(N-k)].
\eeq
\\
We note that the above formulation also directly elucidates the existence of the Hopf insulator~\cite{Hopf_1, Hopf_2,Unal19_PRR, ChernHopf_Yu, Hopf_3}. That is, a two-band system is characterized by the complex Grassmannian $\mathsf{Gr}_{1,2}(\mathbb{C})$, being the Riemann sphere. Considering the third homotopy then coincides with Hopf fibration $S^3 \rightarrow S^2$ \cite{Hopf_1}. 
As a next step, on fixing the orientation, which corresponds to enforcing the Bloch eigenvectors to span oriented frames~\cite{bouhonGeometric2020}, the effective target space of the introduced four-band Hamiltonian extends to an oriented real Grassmanian,\\
\beq{}
    \widetilde{\mathsf{Gr}}_{1,4}(\mathbb{R}) = SO(4)/SO(3) \cong S^3,
\eeq
\\
where we note that a general oriented real Grassmannian is defined as
\\
\beq{}
    \widetilde{\mathsf{Gr}}_{k,N}(\mathbb{R}) = SO(N)/[SO(k) \times SO(N-k)].
\eeq
\\
Hence, in three-dimensions, the Pontryagin index characterizing winding on a three-sphere, as in the high-energy physics of BPST instantons \cite{Zee2010-ZEEQFT, fradkin}, is a natural invariant introduced by the classifying spaces of real four-band Hamiltonians $H(\kv)$ in which one partitions the system into a three-band and single-band subspace. This is consistent with general classification results on real topology \cite{bouhonGeometric2020}, and is equivalent to the elements of third-homotopy groups, independent of the orientability 
\\
\beq{}
    \pi_3 (\widetilde{\mathsf{Gr}}_{1,4}(\mathbb{R})) \cong \pi_3 (S^3) \cong \mathbb{Z},
\eeq
\beq{}
    \pi_3 (\mathsf{Gr}_{1,4}(\mathbb{R})) \cong \pi_3 (\mathbb{RP}^3) \cong \mathbb{Z}.
\eeq
\\
We remark, that such real topology can be viewed as a higher-dimensional analog of orientable and non-orientable three-band Euler Hamiltonians in two-dimensions, which have been realized experimentally in acoustic metamaterials \cite{Jiang_meron,Jiang_2021}. This perspective, as well as the experimental accessibility of three spatial dimensions, offer a platform for realizing the Hamiltonian which we detail in the subsequent.

\subsection{Pontryagin index}
We first identify the Pontryagin index as a $\mathbb{Z}$-valued bulk invariant realized in our settings. The Pontryagin index captures a winding on a three-sphere $S^3$. This winding can be explicitly imposed on a Bloch eigenvector corresponding to the isolated band subspace, or equivalently to the normal bundle $NS^3$. The fourth Bloch band $\nv_4(\kv) \equiv \ket{u_4(\kv)}$ generates the winding of the Hamiltonian, analogously to the third band constituting the frame basis of the two-sphere normal bundle $NS^2$, $\nv_3(\kv) \equiv \ket{u_3(\kv)} = \ket{u_1(\kv)} \times \ket{u_2(\kv)}$ in a two-dimensional Euler insulator with Hamiltonian $H^{\chi}(\kv) = 2 \nv_3(\kv)\otimes \nv^{\text{T}}_3(\kv) - \mathbb{1}_3$ \cite{Eulerdrive, bouhon2019nonabelian, tomas}. The associated higher-dimensional invariant, equating to the Pontryagin index, is given by\\
\beq{Pont_index}
    Q = \frac{1}{2\pi^2} \int_{S^3} \dd^3\kv~
    \varepsilon_{ijkl} (\nv_4)_i \partial_{k_x} (\nv_4)_j \partial_{k_y} (\nv_4)_k \partial_{k_z} (\nv_4)_l 
\eec
\\
where $(\nv_4)_i$ labels the components of the winding vector $\nv_4 $, which can be equivalently expressed in terms of the three other bands as $(\nv_4)_i = \varepsilon_{ijkl}(\nv_1)_j (\nv_2)_k (\nv_3)_l$. This formula is a higher-dimensional ($S^3$ instead of $S^2$) winding analog of the Euler invariant $\chi$ in two-dimensional non-Abelian insulators, that can be deduced from the skyrmion number formula \cite{bouhon2019nonabelian, Eulerdrive}. That is,
\\
\beq{}
    \chi = \frac{1}{2\pi} \int_{S^2} \dd^2\kv~\nv_3 \cdot (\partial_{k_x} \nv_3 \times \partial_{k_y} \nv_3).
\eeq
\\
Moreover, with any vector $\nv_4$, we can associate an $SU(2)$-valued quaternion matrix
\\
\beq{}
    U = (\nv_4)_0 \mathbb{1}_2 + i(\nv_4)_j \sigma_j 
\eec
\\
where Einstein summations are implied and $\sigma_j$ with $j = x,y,z$ correspond to the usual Pauli matrices. We can interpret such a unitary matrix in terms of a non-Abelian $\mathfrak{su}(2)$ connection form $G$, which is \textit{not} the general 
non-Abelian Berry connection commonly used in studying the band topology \cite{bouhon2019wilson}, appears as the connection on the principal $G$-bundle 
for the $SU(2)$ instantons \cite{fradkin}, and is defined as 
\\
\beq{}
    G = U^{-1} \dd U,
\eeq
\\
with  associated non-Abelian curvature $\mathcal{F}$,
\\
\beq{}
    \mathcal{F} = \dd G + G \wedge G.
\eeq
\\
In these terms, the Pontryagin index can be written as
\\
\beq{eq:Tr3}
    Q = \frac{1}{24\pi^2} \int_{S^3} \text{Tr}~(U^{-1} \dd U)^3 = \frac{1}{24\pi^2} \int_{S^3} \text{Tr}~G^3.
\eeq
\\
Interestingly, one may show that for four-band phases split into one- and three-band subspaces, the Pontryagin index, that is the bulk invariant corresponding to the winding number of the isolated Bloch vector of the proposed models, can be constructed in terms of the non-Abelian Berry connection.

The non-Abelian Berry connection elements are defined as
\\
\beq{}
    A^{\alpha}_{ij} = \bra{u_i} \partial_{k_\alpha} \ket{u_j}
\eec
\\
with band indices $i,j = 1, 2, 3, 4$ and momentum indices $\alpha = x, y, z$. As detailed in Appendix \ref{App::A}, one may show that 
\\
\begin{eqnarray}\label{QinConnection}
Q &=& \frac{1}{2\pi^2} \int_{T^3} \dd^3\kv~ \vec{A}_{41} \cdot (\vec{A}_{42} \times \vec{A}_{43}) \\ \nonumber
&\equiv& \frac{1}{2\pi^2} \int_{T^3} \dd^3\kv~ \Big[\vec{A}_{41}, \vec{A}_{42}, \vec{A}_{43} \Big], 
\end{eqnarray}
\\
where the $\mathfrak{so}(4)$ connection elements connect the occupied and unoccupied band subspaces, analogously to the other invariants characterizing non-Abelian phases \cite{Jiang_2021,lim2023real}. For example, in two spatial dimensions, the three-band Euler invariant can be rewritten as
\\
\beq{}
    \chi = \frac{1}{2\pi}\int_{T^2} \dd^2\kv~ \varepsilon_{\alpha\beta} A^{\alpha}_{31} A^{\beta}_{32} 
\eec
\\
where $\varepsilon_{\alpha\beta}$ is a $2 \times 2$ real antisymmetric matrix with unit determinant. 

From the perspective of Eq.~\eqref{QinConnection}, the invariant is viewed as an integral of the connection field volume \text{three-form} obtained from the connection vectors associated with the Bloch bundle over the base hypertorus $T^3$.

\subsection{Relation to Pontryagin class}

As a next step, we elaborate on the connections of the bulk Pontryagin index to the closely related higher-dimensional characteristic class, the Pontryagin class. The Pontryagin class characterizing four-dimensional topological phases with reality condition, can be written in terms of $SO(4)$-valued non-Abelian Berry curvature $F^{\alpha \beta}_{ij} = \partial_\alpha A^\beta_{ij} - \partial_\beta A^\alpha_{ij} + [A^\alpha,A^\beta]_{ij}$ as \cite{bouhon2023second}
\\
\beq{eq:P1}
P_1 = \frac{1}{8\pi^2} \int_{T^4} \dd^4 k ~ \varepsilon_{\alpha \beta \gamma \delta} F^{\alpha \beta}_{ij} F^{\gamma \delta}_{ij},
\eeq
\\
where the integration measure includes all momenta $k_\alpha$ with $\alpha = 1,2,3,4$ present, or more generally four parameters of a parameter space, e.g. three momenta and additional parameter $t$, in the context of this work. 


Upon relaxing the reality condition, a complexification of the real Bloch bundle allows us to redefine the associated characteristic class as a second Chern number 
\\
\beq{}
C_2 = \frac{1}{8\pi^2} \int_{T^4} \dd^4k~ \varepsilon_{\alpha \beta \gamma \delta} \tilde{F}^{\alpha \beta}_{ij} \tilde{F}^{\gamma \delta}_{ij}
\eec
\\
where $\tilde{F}^{\alpha \beta}$ is the non-Abelian Berry curvature over the complexified bundle, traced over the occupied bands $i,j$. This is consistent with the relation between characteristic classes~\cite{Nakahara}
\\
\beq{}
    p_k(E) = (-1)^k c_{2k}(E \oplus iE)
\eec
\\
where $E$ denotes the total space of a real Bloch bundle $\mathcal{B}$ and $E \oplus iE$ is its complexification for any  arbitrary positive integer $k$. We stress that the non-triviality of the first Pontryagin class demands reality of the bundle, hence the necessity for enforcing a symmetry such as $\mathcal{PT}$. Additionally, the Pontryagin class is only defined for vector bundles of dimension $4k$, as in terms of cohomology rings $p_k(E) \in H^{4k}(S^{4k}, \mathbb{Z}) \cong \mathbb{Z}$,  meaning that the lowest-dimensional Pontryagin insulator requires four dimensions. However, a four-dimensional Pontryagin insulator requires at least six bands for non-triviality of the invariant \cite{bouhon2023second}. Hence, it cannot be dimensionally reduced to our three-dimensional model in the manner in which an axion insulator can be seen as a descendant of a second Chern insulator \cite{Qi_2008}. We may however construct an artificial setup to relate to the Pontryagin class in four dimensions, without inducing extra bands. For this, we begin by dimensionally extending the eigenvectors to generate a new Hamiltonian $H(\kv,t) = 2 \ket{u_4(\kv, t)} \bra{u_4(\kv, t)} - \mathbb{1}_4$. We demand $t$ to be a real parametrization in range $t = 0 \rightarrow t = \infty$. Setting, 
\\
\beq{}
    \ket{u_4(\kv, t)} = \sqrt{f(t)} \ket{u_4(\kv)}
\eec
\\
with a smooth function $f(t) = \frac{t^2}{t^2+1}$, we may induce an extended non-Abelian connection, which is specifically of $\mathfrak{su}(2)$ type exactly in the $t = \infty$ limit, with~\cite{fradkin} 
\\
\beq{}
    G = f(t) U^{-1} \dd U.
\eeq
\\
In this picture, physically the state $\ket{u_4(\kv, t)}$ does not change as soon as $t \neq 0$. This can be understood by recognizing the transformation as a time-dependent rescaling, which can be removed by normalization, as long as the vector is non-vanishing ($t \neq 0$). We stress that contrary to the non-Abelian Berry connection, the extended connection, here used only to establish a link to Pontryagin characteristic class, does not require normalization of the Bloch states $\ket{u_4(\kv, t)}$, contrary to the states $\ket{u_4(\kv)}$ used for the quaternion construction of the $SU(2)$ matrices $U$, which necessarily need to be normalized to achieve the unitarity of $U$. Mapping to the associated curvature two-form upon taking an exterior derivative, the connection yields
\\
\beq{}
Q = \frac{1}{8\pi^2} \int_{D_4} \text{Tr} \mathcal{F} \wedge \mathcal{F} = P_1,
\eeq
\\
defined on an open disk $D_4$, which is bounded by a three-sphere $S^3$ parametrized with crystal momentum. Here, the fourth $k$ coordinate $t$ can be interpreted as a tuning parameter that allows for the expansion of the sphere~\cite{fradkin}.

The above thus shows that one may establish a connection between the Pontryagin index invariant present in the proposed three-dimensional models and, on dimensional extension, one of the four characteristic classes capturing the topology of vector bundles corresponding to band structures.

\section{Models}\label{sec::III}
Within this section we formulate concrete models allowing us to induce non-trivial Pontryagin index topology. We begin by formulating a minimal model, after which a systematic framework is introduced to generate models exhibiting arbitrary values of the invariant. 

\subsection{Flat band limit}
An effective approach to formulate a minimal model is to appeal to the flat band limit to capture the real topology of topological phases with non-trivial Pontryagin class. We use a minimal construction, which explicitly induces a winding on $S^3$ by the fourth-band~\cite{Eulerdrive,bouhon2019nonabelian,bouhon2023second,tomas}, while keeping three other bands separated from the fourth by a gap, namely, 
\\
\beq{Pont3D_H}
    H^{Q}(\kv) = 2 \nv_4(\kv)~\otimes~\nv^{\text{T}}_4(\kv) - \mathbb{1}_4.
\eeq
\\
The non-trivial Pontryagin index corresponding to the above system~\eqref{Pont3D_H} is then simply given as
\\
\beq{eq:explicitmodel}
    \nv_4(\kv) = \frac{1}{\mathcal{N}}
    \begin{pmatrix}
     \sin{p_x k_x} \\
     \sin{p_y k_y} \\
     \sin{p_z k_z} \\
     m - \sum_{i = x,y,z} \cos{p_i k_i} \\
    \end{pmatrix}.
\eeq
\\
In the above, $\mathcal{N}$ represents a normalization factor and the parameters $p_x, p_y, p_z \in \mathbb{Z}$ are introduced to allow for arbitrary Pontryagin indices ${Q = 2 p_x p_y p_z}$ and $Q = p_x p_y p_z$ on changing $m$. We reiterate that viewed as winding on $S^3$, any Pontryagin index corresponds to a member of the third homotopy group $\pi_3 ( S^3 ) \cong \mathbb{Z}$. 

Importantly, the possibility of an arbitrarily high associated winding invariant $Q$, namely Pontryagin index, definitionally implies the presence of a $\mathbb{Z}$-type invariant, which we also verify further with the $\mathbb{Z}$-type bulk-boundary correspondence (Sec.~\ref{sec::IV}).

\subsection{Pl\"ucker embedding}
Alternatively, one may also generally construct a Hamiltonian of any Pontryagin index $Q$ through the Pl\"ucker embedding approach~\cite{bouhonGeometric2020, braidingtwo, bouhon2023quantum}.
Previously, this framework was used in the context of two-dimensional Euler insulators \cite{bouhonGeometric2020} and second Euler insulators \cite{bouhon2023second}, which are close relatives of the truly four-dimensional Pontryagin insulator \cite{bouhon2023second}. In brief, the embedding encapsulates equipping elements of the classifying Grassmannians with multi-vectors, which form the basis for matrix construction of the Hamiltonian \cite{bouhonGeometric2020}. We start with a flattened Hamiltonian and a Bloch band matrix $R(\kv)$\\
\beq{}
    H^Q(\kv) = R(\kv)
    \begin{pmatrix}
        -1 & 0 & 0 & 0 \\
        0 & 1 & 0 & 0 \\
        0 & 0 & 1 & 0 \\
        0 & 0 & 0 & 1 \\
    \end{pmatrix} R(\kv)^{\text{T}}.
\eeq
\\
The matrix $R(\kv) \in SO(4)$ is generated via a parametrization with three angles $(\phi, \psi, \theta)$, where those are provided by maps
\\
\beq{}
    \psi(\kv) = \pi~\text{max}\{|k_x|, |k_y|, |k_z|\} 
\eec
\\
\beq{}
    \theta(\kv) = \cos^{-1}(k_z/\sqrt{k^2_x + k^2_y + k^2_z})
\eec
\\
\beq{}
    \phi(\kv) = Q \tan^{-1}(k_x/k_y)
\eec
\\
for any $k$-point in the BZ isomorphic to a three-torus $T^3$. Intuitively, the map assigns angular hyperspherical coordinates $(\phi, \psi, \theta)$ to different patches of $k$-points in the BZ, while inducing the angular winding introduced by $Q$. The parametrization of the Hamiltonian is explicitly provided by a map
\\
\beq{}
    R(\kv) =  \text{e}^{i\theta\Gamma_{20}} \text{e}^{i\phi\Gamma_{21}}
    \text{e}^{i\psi\Gamma_{12}}
\eec
\\
with corresponding matrices $\Gamma_{ij} = \sigma_i~\otimes_K\sigma_j$, where $\otimes_K$ denotes a Kronecker product. We stress that Fourier transforming the components of $H^Q(\kv)$ would in principle yield long-range hoppings, which can be truncated to finitely ranged neighbor hoppings without changing the topology as long as the gap is not closed upon truncation, which we effectively corroborate at each stage of the analysis~\cite{bouhon2019wilson, bouhonGeometric2020, bouhon2023quantum}.

The Bloch vectors, which can be identified with columns of $R(\kv)$ as seen through the spectral decomposition, can be projected on $S^3$ as a base space to form a vector bundle, such that the three-band subspace spans its tangent bundle $TS^3$, while the last band constitutes the normal bundle $NS^3$.
The maps introduced above generate the winding in the bundle and the Pontryagin index can be viewed as the winding of the fiber across the bundle. 

\section{Physical manifestations of topology}\label{sec::IV}

Having introduced the mathematical aspects of the Pontryagin topology and, accordingly, model realizations, we proceed to the main results corresponding to the manifestations of non-trivial hyperspherical windings and underlying non-Abelian structures, as captured by the well-established $S^3 \cong SU(2)$ isomorphism.

\subsection{Bulk-boundary correspondence}

\begin{figure}
\centering
  \includegraphics[width=\columnwidth]{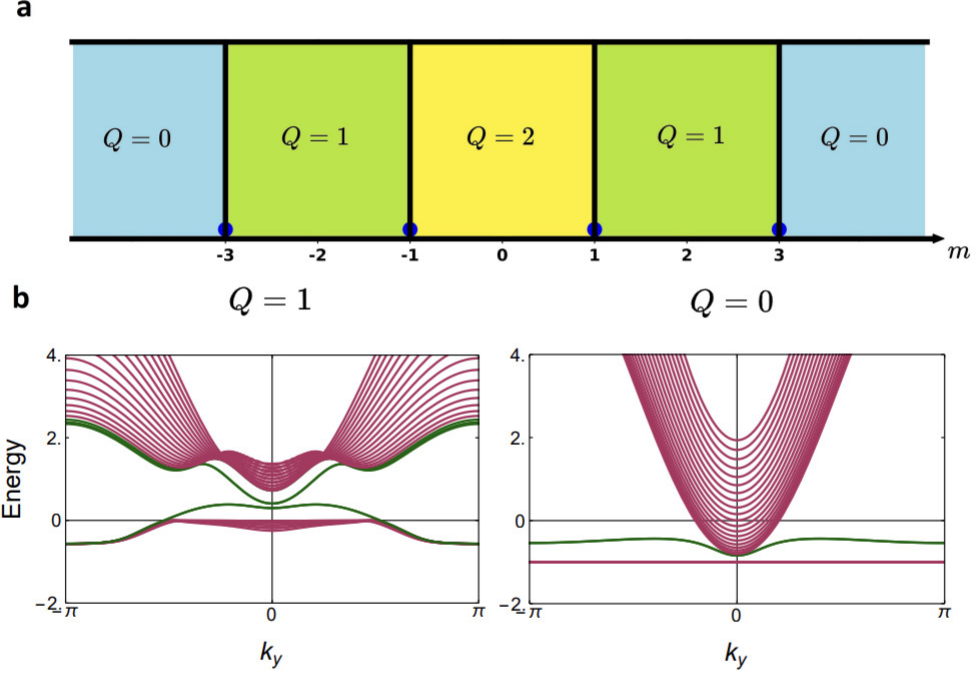}
  \caption{\small 
  (\vec{a}) Phase diagram of the minimal model with $p_x = p_y = p_z = 1$ supporting non-trivial Pontryagin indices $Q$. The mass term controls the $\mathbb{Z}$-valued invariant, similarly to the two-dimensional models with Euler topology~\cite{Eulerdrive}.
  (\vec{b}) Projection of bulk bands and edge states along a one-dimensional section of a three-dimensional insulator with non-trivial Pontryagin index $Q = 1$ set in the minimal model, as compared with $Q = 0$.}
\label{fig:Plot1}
\end{figure}

We begin by investigating the spectrum and bulk-boundary correspondence in the minimal model with $(p_x,p_y,p_z)=(1,1,1)$ set as in Eq.~\eqref{eq:explicitmodel}, with the occupied three-band subspace, and unoccupied band inducing the non-trivial Pontryagin index. The projected band structure of the insulator with $Q = 1$, which includes the projections of the boundary states, is shown in Fig.~\ref{fig:Plot1}. 

We observe that despite the presence of edge modes attainable at any energy within the bulk gap, these do not need to connect valence and conduction bands. Interestingly, in the context of the other $\mathcal{PT}$-symmetric phases, the modes with a similar property were reported in two-dimensional meronic Euler insulators, also characterized by non-Abelian topology~\cite{Jiang_meron}. In particular, we notice that the obtained one-dimensional projection of states, retaining only $k_y$ as a good quantum number labeling the states (see Fig.~\ref{fig:Plot1}), is reminiscent of the one-dimensional projected spectrum obtained from the two-dimensional Euler Hamiltonians. As will be explored further in the following, the connection to the Euler phases becomes stronger, on recognizing that the bulk Hamiltonian with $Q \neq 0$ can support surface bands with a {\it surface Euler invariant}, subject to the constraint that the boundary satisfies the reality condition~\cite{bouhon2019nonabelian,bouhonGeometric2020}.

We further note that upon taking $m = 2$, the edge states can be separated by an energy gap, if $\mathcal{PT}$-symmetry is broken on the surface.~Correspondingly, there exist surface bands which can acquire a dispersion reminiscent of the gapped pockets in the two-dimensional edge spectrum found in three-dimensional axion insulators (Fig.~\ref{fig:Plot1})~\cite{Varnava_2018}. 
Moreover, the edge state branches can be disconnected from the projected bulk with purely $\mathcal{PT}$-symmetric perturbations. We stress again that for the non-Abelian phases studied in this work, the symmetry is spinless $(\mathcal{PT})^2 = 1$. To contrast the introduced $\mathcal{PT}$-symmetric phases with the axion insulators further, we may also compare the bulk states, rather than the boundary eigenstates. The bulk of axion insulators is characterized by the $\theta$ angle, proportional to the magnetoelectric polarizability, and given by an integral of the Chern-Simons three-form
\\
\beq{}
    \theta = \frac{1}{4\pi} \int_{T^3} \operatorname{Tr}\left[ \dd A \wedge A + \frac{2}{3} A \wedge A \wedge A \right].
\eeq
\\
Here, the trace is evaluated over occupied states. Interestingly, on direct evaluation of $\theta$ for arbitrary $Q \in \mathbb{Z}$, we find $\theta_v = 4\pi Q$ in the valence subspace of three bulk bands, with $\theta_c = 0$ trivially in the single conduction band satisfying the reality condition. However, under the $3 \oplus 1$ band partitioning, a gauge transformation on the occupied states acts as $g \in U(3)$ in the degenerate limit, obtaining a unitarily-equivalent set of single-particle eigenstates, which transforms the $\theta_v$ as~\cite{vanderbilt2018berry}
\\
\beq{}
    \theta_v \rightarrow \theta_v + \frac{1}{12\pi} \int_{T^3} \dd^3\kv~ \varepsilon^{pqr} \text{Tr} \big[ (g^{-1} \partial_p g) (g^{-1} \partial_q g) (g^{-1} \partial_r g) \big].
\eeq
\\
Such a gauge term can be identified with the $\mathbb{Z}$-valued generator of the cohomology $H^3(U(3), \mathbb{Z}) \cong \mathbb{Z}$, namely $\frac{1}{24\pi^2} \int_{S^3} \text{Tr} (g^{-1} \dd g)^3$ \cite{PhysRevLett.110.067205}, yielding a gauge ambiguity of $\theta_v \rightarrow \theta_v + 2\pi n$, where $n$ is integer. Hence, $\theta_v$ can only be defined $\text{mod}~2\pi$. Therefore, for any introduced $Q$ we find, ${\theta_v = 4\pi Q \rightarrow 0~(\text{mod}~2\pi)}$.

As follows from the known bulk-boundary correspondence for the three-dimensional topological insulators~\cite{Qi_2008}, the triviality of the bulk $\theta$ allows the surface anomalous Hall conductivity to vanish. Indeed, we consistently find that the total surface Chern number ($C_{s,\text{tot}}$) vanishes, which is consistent with the surface Chern theorem~\cite{PhysRevB.95.075137}. To evaluate $C_{s,\text{tot}}$, we compute the surface bands $\ket{u^s_{n}(k_x,k_y)}$ over the reduced BZ ($\text{rBZ} \cong T^2$), as well as the associated Wannier ladders (see Appendix~\ref{App::B}). Explicitly, the total surface Chern number ($C_{s,\text{tot}}$) is defined as
\beq{}
    C_{s,\text{tot}} = \frac{1}{2\pi} \int_{\text{rBZ}} \dd^2\kv~ \sum_n \Omega^n_{xy,s}. 
\eeq
Here, $\Omega^n_{xy,s} \equiv i[\bra{\partial_{k_x} u^s_{n}(k_x,k_y)}\ket{\partial_{k_y} u^s_{n}(k_x,k_y)} - \text{c.c.} ]$ is the Abelian Berry curvature of the surface band, which individually integrates to $C^n_s$. Here, $n$ runs over all bands with the hybrid Wannier centers $\bar{w}_z(k_x,k_y)$ self-contained in the surface layers. The topological nature of the surface states for non-vanishing $Q$ is captured by the $2Q$-fold winding of $\bar{w}_z(k_x,k_y)$, as can be observed in the Wannier ladders along arbitrary quasimomenta $k_x$ and $k_y$, see Appendix~\ref{App::B}.

Having introduced $C_{s,\text{tot}}$, we recognize that the surface Chern theorem gives $\theta_c+\theta_v = 2\pi C_{s,\text{tot}}$, as also shown in Ref.~\cite{lim2023real} in the context of $\mathcal{PT}$-symmetric real Hopf insulators (see also Sec.~\ref{sec::V}). Correspondingly, we obtain $\theta_v = 4 \pi Q = 2\pi C_{s,\text{tot}} = 0~\text{mod}~2\pi$. Therefore, $C_{s,\text{tot}} = 0$, on inserting the previously recognized $\theta_c = 0$, and on including the modular character of $\theta_v$ ($\text{mod}~2\pi$) for the three-band occupied valence subspace. This finding is distinct from real Hopf insulators
with bulk-boundary correspondence captured by a pair of Hopf indices resulting in surface Chern numbers equal and opposite on the opposite boundaries; a configuration which also manifestly satisfies $\mathcal{PT}$-symmetry.

However, while $C_{s,\text{tot}} = 0$ for any Pontryagin index~$Q$, we crucially find that the boundaries host topologically nontrivial pairs of surface bands. Namely, we calculate the surface Wilson loop spectrum, obtaining nontrivial Wilson loop eigenvalue windings of $\pm 2Q$. Consistently with the surface Chern theorem, this configuration yields $C_{s,\text{tot}} = 2Q - 2Q = 0$. Furthermore, as long as the surface obeys $\mathcal{C}_2\mathcal{T}$-symmetry, the pair of the surface bands can be characterized with a surface Euler topology under the reality condition imposed on the surface~\cite{bouhon2019nonabelian,bouhonGeometric2020}. The surface Euler invariant ($\chi_s$) reads:
\beq{}
    \chi_s = \frac{1}{2\pi}\int_{\text{rBZ}} \dd^2\kv~ \text{Eu}_{xy,s}, 
\eeq
with integrand, ${\text{Eu}_{xy,s} \equiv \bra{\partial_{k_x} u^s_{n}(k_x,k_y)}\ket{\partial_{k_y} u^s_{n+1}(k_x,k_y)}} \\ - \bra{\partial_{k_y} u^s_{n}(k_x,k_y)}\ket{\partial_{k_x} u^s_{n+1}(k_x,k_y)}$, the surface Euler curvature in surface bands $n,n+1$. Consistently with the obtained surface Wilson loop windings, this amounts to $\chi_s = 2Q$. While the numerical validation for the discussed correspondence is provided in Appendix~\ref{App::B}, we leave the further study of the emergence of such exotic bulk-boundary correspondence for future work.

Finally, we remark that for any general non-vanishing value of $Q$, the bulk-boundary correspondence is reflected by the topological protection of the edge states accompanying the topologically non-trivial bulk ($Q \neq 0$). In particular, it is manifested by their robustness against disorder, as numerically validated in Sec.~\ref{sec::VI}.

\subsection{Topological phase transitions}

We can furthermore construct a phase diagram from the minimal model introduced above. When $p_x = p_y = p_z = 1$, we observe two topological phase transitions at $m = \pm 1$ and $m = \pm 3$. Correspondingly, the Pontryagin index changes from $Q = 2$ to $Q = 1$ and to $Q = 0$, as shown in Fig~\ref{fig:Plot1}. We notice that this finding is different from the two-dimensional finding in orientable Euler insulators~\cite{Bouhon2020}, where $\chi = 2N$ with $N  \in \mathbb{Z}$, admitting only a direct transition with $\Delta \chi = \pm 2$, e.g. from $\chi = 2$ to $\chi = 0$ alongside a removal of Euler nodes, on closing the principal gap. Interestingly, as the trivialization occurs on closing the principal gap in three-dimensional Pontryagin models, the nodal structure also disappears, as we detail in the subsequent.

The reduction of Pontryagin index through topological phase transitions can be viewed in terms of the trivialization of subdimensional non-Abelian Wilson loop windings (see also Sec.~\ref{sec::IV}C, Fig.~\ref{fig:WilsonLoop}) over two-dimensional sections of BZ at the high-symmetry planes $k_z = 0$ and ${k_z = \pi}$. To understand this further, we recognize that the momentum-space construction of the Hamiltonian resembles the construction of a strong three-dimensional, spinful, $\mathbb{Z}_2$ insulator, where the strong invariant is induced on appropriately coupling two-dimensional $\mathbb{Z}_2$ quantum spin Hall insulator models at $k_z = 0$ and $k_z = \pi$. Here, the induced strong invariant is the Pontryagin index, which is $\mathbb{Z}$ valued, as reflected by the following construction from Euler insulators hosting $\mathbb{Z}$ invariants. Namely, the Hamiltonian spanning from $k_z = 0$ to $k_z = \pi$ momentum-space planes, as defined in Eq.~\eqref{eq:explicitmodel}, can be thought of as equivalent to a family of individual subdimensional non-Abelian 2D phases \cite{Eulerdrive}; (i)~initially consisting of three bands in total: two occupied Euler bands, and one unoccupied band given by $H^{\chi}(\kv)$; (ii)~modified by an inclusion of a trivial occupied band accompanying the two-band subspace; (iii)~with a fixed mass term set by the $k_z$ ($m^* = m - \cos k_z$). We note that the other subdimensional relations in terms of two-dimensional fragile, though symmetry-indicated rather than symmetry-indicator-free~\cite{bouhonGeometric2020} topology, were established in the context of axion insulators with $\theta = \pi$ \cite{Axion3, Bouhon2020, Lange2021}. 

\begin{figure}
\centering
\includegraphics[width=\columnwidth]{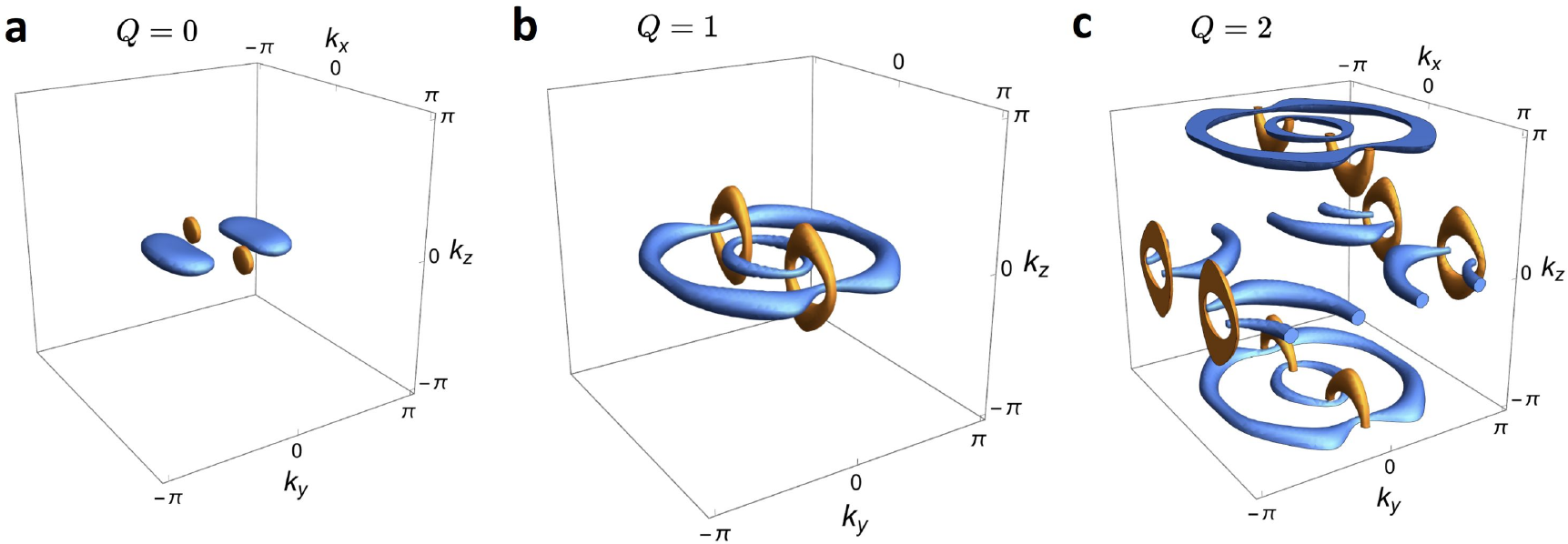}
  \caption{\small Nodal structures over the three-dimensional Brillouin zone for the minimal model of the Hamiltonian provided in Eq.~\eqref{Pont3D_H}. Nodes between bands $\ket{u_1}$ and $\ket{u_2}$ were denoted in blue, while band crossings between bands $\ket{u_2}$ and $\ket{u_3}$ were denoted in orange.  (\vec{a}) $Q=0$ with trivial nodal structure and vanishing linking numbers. (\vec{b}) $Q=1$, with two nodal rings present. (\vec{c}) $Q=2$, with four linked nodal rings. We show that the structures are not protected by the Pontryagin index $Q$, but are easily realized in the context of the presented models.}
  \label{fig:examplenodes}
\end{figure}

In the next section, after discussing the nodal structures that naturally arise in the three-band subspaces of our models, we further propose topological phase transitions to more general multigap flag limit phases with four isolated bands, classified by an oriented flag variety 
\\
\beq{}
    \widetilde{\mathsf{Fl}}_{1,1,1,1}(\mathbb{R}) = SO(4)\cong \frac{S^3 \times S^3}{\mathbb{Z}_2}
\eec
\\
that thereby serve as an unambiguous reference to further interpret the outlined topological structures. Here we note that a general oriented flag manifold is defined as
\\
\beq{}
    \widetilde{\mathsf{Fl}}_{p_1,..,,p_N}(\mathbb{R}) = SO(4)/[SO(p_1) \times ... \times SO(p_N)].
\eeq
\\
Such multigap flag phases where all bands are fully partitioned can be accessed on addition of a proper term to the Hamiltonian with Pontryagin index $Q$, or smooth reparametrization of the diagonal matrix in the embedding construction, in both cases necessarily removing any band crossings to enter the full multigap regime. We will moreover see that the connection to the flag limit (Sec.~\ref{sec::V}) will provide for an unambiguous reference to fully elucidate the above topological features. That is, we will show that such topological transitions can be tracked with the bulk index, as well as the distinct bulk-edge correspondence for various types and values of the bulk invariants (Sec.~\ref{sec::VI}). Before turning to these aspects, we, however, first comment on the naturally emerging linking structures in our models (see Fig.~\ref{fig:examplenodes}).

\subsection{Nodal structures}

In this section, we discuss the nodal structures present in the three-band subspaces within the band structures of the introduced models. Such a characterization is necessary to further understand the bulk invariant $Q$ induced by $\pi_3(\mathsf{Gr}_{1,4}) \cong \pi_3(S^3) \cong \mathbb{Z}$, central to this work, given that the nodal topology is supported by the tangent bundle $TS^3$ of the classifying space, which is parallelizable \cite{Hatcher_2}, in contrast to the winding of the isolated band in the normal bundle $NS^3$ inducing the invariant. The parallelizability of the tangent bundle of $S^3$ implies that the nodal structure cannot be responsible for the value of $Q$, the manifestation of which we demonstrate by explicit unbraiding constructions outlined in this section. We stress that these nodal links appear very naturally in our models, as shown in Fig.~\ref{fig:examplenodes}, and can be unbraided~\cite{Tiwari} quite efficiently to further enter the full flag limits introduced in the previous subsection. As a result, we empirically observe that the presented model setting provides an excellent platform to accomplish braiding in rather simple four-band models that should appeal to metamaterials settings, see also Section \ref{sec::VII}. Here, we elaborate on the non-Abelian charges carried by the nodal structures present in the models, and on the unbraiding necessary to access the reference flag limits further discussed in the next section. We include explicit parametrizations of the corresponding braiding Hamiltonians in the Appendix~\ref{App::C}.

\begin{figure*}[t!]
\centering
  \includegraphics[width=\linewidth]{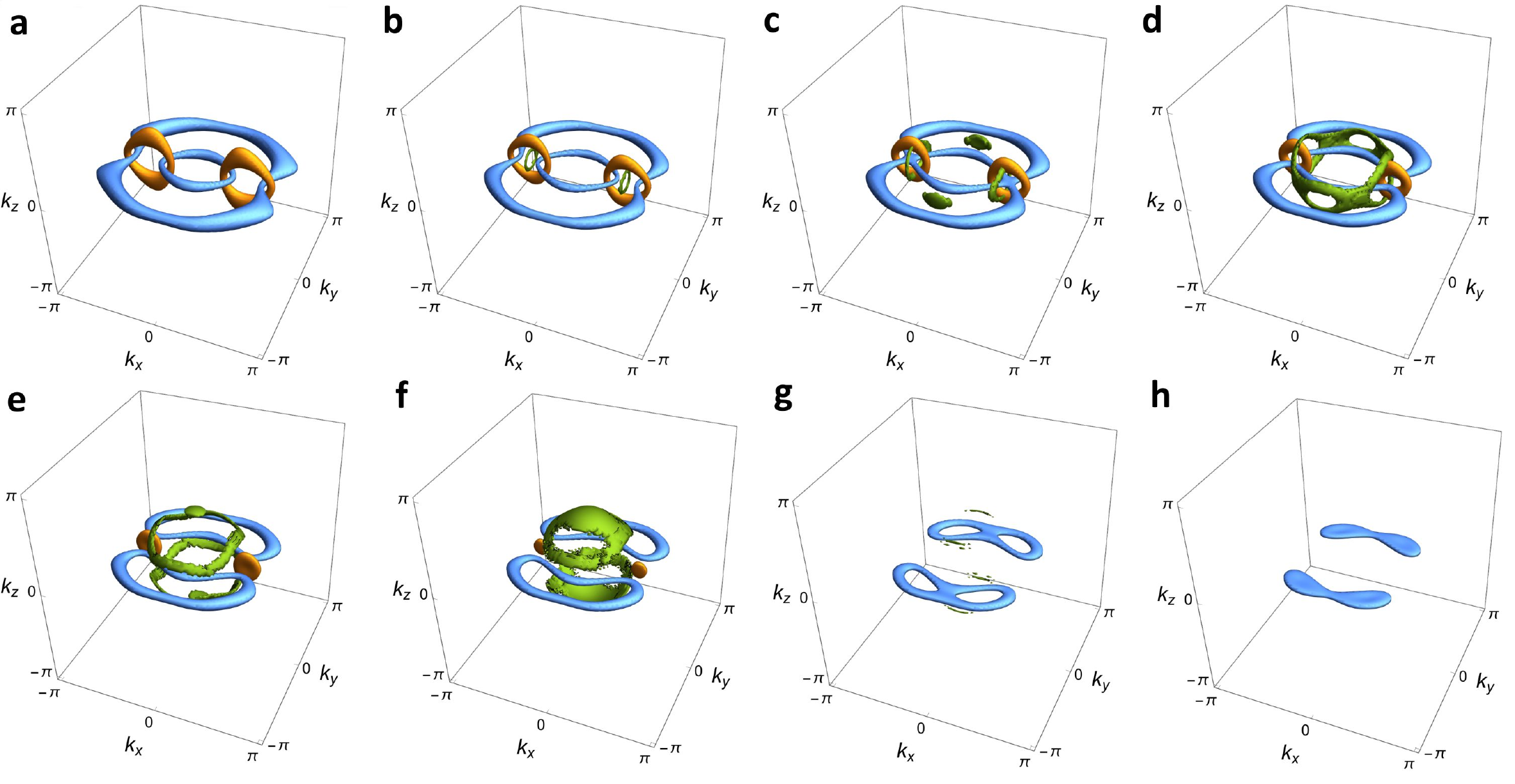}
  \caption{\small Explicit unlinking of nodal structure through the creation of adjacent nodal rings on closing a gap, as parametrized in Appendix~\ref{App::C}. The blue and orange rings show nodal lines between the bands ($\ket{u_1}$--$\ket{u_2}$, and $\ket{u_2}$--$\ket{u_3}$, correspondingly) in the occupied three-band subspace and the green rings show nodal lines between the highest occupied and unoccupied bands. 
  (\vec{a}) initial nodal links, (\vec{b})--(\vec{c}) creation of additional rings, (\vec{d}) connecting rings, (\vec{e})--(\vec{f}) splitting and disentangling green and blue rings, (\vec{g}) removal of green rings, (\vec{h}) contraction and annihilation of blue rings to enter the gapped flag limit.
  Numerically, the unbraiding on closing the principal gap can be achieved by adding a diagonal mass term to the Hamiltonian, similarly to the two-dimensional Euler insulators and semimetals, where an effective mass to unbraid the nodes can also be realized by adding onsite disorder~\cite{jankowski2023disorderinduced}.}
  \label{fig:unbraiding_closed}
\end{figure*}

To characterize the nodal topology, we start by noticing that the frames $\{ \ket{u_1}, \ket{u_2}, \ket{u_3}, \ket{u_4} \}$ constituting vierbeins at any $k$-point can acquire an accumulated angle on being parallel-transported around any node due to band touching. These are captured by the first homotopy group of an unoriented flag variety $\mathsf{Fl}_{1,1,1,1}$ with $O(4)$ representing general rotations of the vierbein and each $\mathbb{Z}_2$ capturing the gauge freedom of single Bloch vector, $\ket{u_i} \rightarrow -\ket{u_i}$, enforced by the real symmetry. The according fundamental group is
\\
\beq{}
    \pi_1(\mathsf{Fl}_{1,1,1,1}) = \pi_1\Big(\frac{O(4)}{\mathbb{Z}_2^4}\Big) \cong \bar{P}_3,
\eeq
\\
where $\bar{P}_3$ is the Salingaros vee group of Clifford algebra $Cl_{0,3}$, which is a non-Abelian group of rank 16, with ten conjugacy classes. Therefore, a quaternion group is a subgroup of such group, implying its non-Abelian character. The Salingaros vee group is obtained from $Cl_{0,3}$ by first defining a basis for the Clifford algebra as 
\begin{equation}
\mathcal{B} = \{1, e_1, e_2, e_3, e_1 e_2, e_2 e_3, e_1 e_3, e_1 e_2 e_3\} \equiv \{e_\textbf{i}\},    
\end{equation}
where the set $\{e_1, e_2, e_3\}$ generates the algebra -- this fact is important when assigning the non-Abelian charges to nodes in the energy spectrum. The vee group is then defined as
\beq{}
\mathsf{G} = (\{ \pm e_\textbf{i}~ | ~e_\textbf{i}  \in \mathcal{B}\}, \times),
\eeq
where $\times$ represents the Clifford algebra multiplication.

There exists a ring isomorphism between $Cl_{0,3}$ and the splitbiquaternions, which are a type of hypercomplex number based on the quaternions. While quaternions are of the form $w + x \textbf{i} + y \textbf{j} + z \textbf{k}$ and have real coefficients $\{w, x, y, z\}$, biquaternions have complex coefficients multiplying the imaginary units, lifting the number of real dimensions from four to eight. The split-biquaternions are then obtained by having the coefficients be split-complex numbers, which are of the form $z = x + \textbf{i} y$, with $\textbf{i}^2 = 1$ rather than $\textbf{i}^2 = -1$. The split biquaternions are also isomorphic to $\mathbb{H} \oplus \mathbb{H}$, where $\mathbb{H}$ are the quaternions. This shows how the quaternion charges of $2 \oplus 1$ band models~\cite{Jiang_2021} can be found as a subgroup of the charges in $3 \oplus 1$ band ones.

The ring isomorphism also implies a group isomorphism between the vee group of $Cl_{0,3}$ and the group of split-biquaternions, thus the charges can be assigned as shown in the Appendix \ref{App::D}. As the nodal topology of four-band non-Abelian insulators with real topology requires non-triviality of the first homotopy group of corresponding flag variety, a similar classification was also achieved in one-dimensional and two-dimensional phases \cite{Jiang_2021}, which however cannot realize the nodal ring structures and associated unbraiding, requiring three spatial dimensions, as discussed in this work.

We find that by construction of the minimal models introduced in Eq.~\eqref{Pont3D_H}, the linking number of \textit{all} nodal rings corresponding to nodes in different gaps is equal to the Pontryagin index $Q$ therein. The topological phase transitions introduced in the previous sections cause the disappearance of the nodal structures through an associated unbraiding, as shown in Fig.~\ref{fig:unbraiding_closed}. This process involves subsequently: creation of additional nodal rings on gap closure, unbraiding which flips the nodal charges, reconnecting the rings, and finally contracting them. On unbraiding, which allows unlinking the structure on flipping the charges, the contraction can remove the nodes, ultimately gapping out the phase.  We give an explicit parametrization of this process in Appendix~\ref{App::C}.

\begin{figure*}[t!]
\centering
\includegraphics[width=\linewidth]{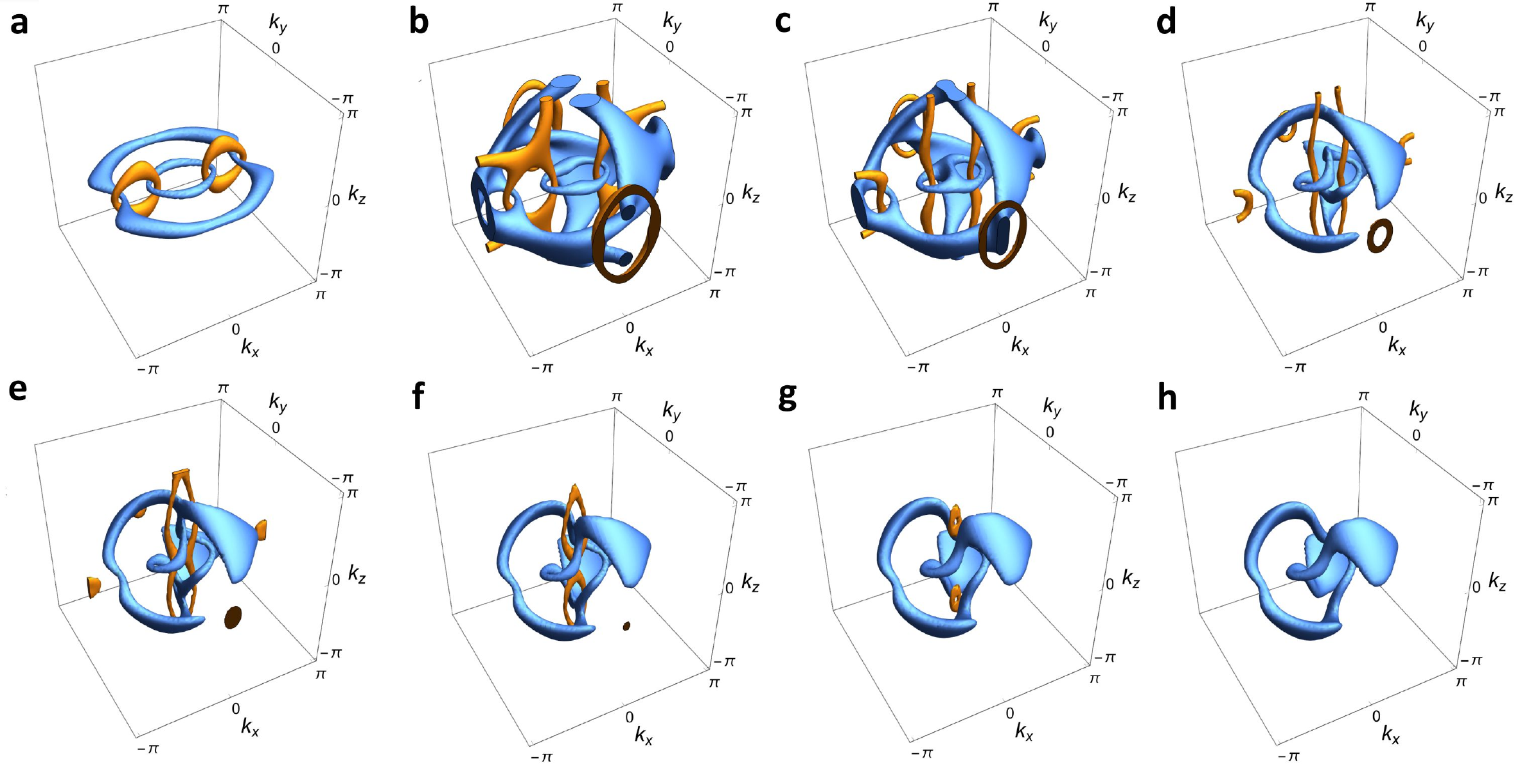}
  \caption{\small Visualization of the process of unbraiding of nodal structures without closing the principal gap in a model with $Q = 1$. The blue and orange rings show nodal lines in the occupied three-band subspace, as in Fig.~\ref{fig:unbraiding_closed}. (\vec{a}) initial nodal links, (\vec{b}) creation of additional rings, (\vec{c}) connecting rings, (\vec{d})--(\vec{e}) splitting and disentangling rings, (\vec{f})--(\vec{g}) closure and removal of orange rings, (\vec{h}) leftover blue ring, which can be contracted and removed, obtaining the fully gapped flag limit. The unbraiding was obtained on adding a parametrized dispersive band-splitting term to the Hamiltonian, contrary to the previously introduced unbraiding with a diagonal term closing the principal gap; see also Appendix~\ref{App::C}.}
  \label{fig:unbraiding_open}
\end{figure*}

However, as we explain, the unbraiding of nodal ring structures does not necessarily require closing the principal gap (the gap between the three-band and single-band subspace), see Fig.~\ref{fig:unbraiding_open}, which is supported by the fact that a tangent bundle of a three-sphere $TS^3$ hosting three-band subspace of our models is parallelizable. This property admits a removal of singularities in the tangent bundle due to the nodes, without accessing the fourth band from the normal bundle $NS^3$ by the principal gap-closing degeneracies. We show by explicit construction that the presence of split-biquaternion nodes is not intrinsically due to the non-triviality of Pontryagin index as a bulk invariant, which would be equivalent to the necessity of closing a gap for the removal and unbraiding of nodal structure $only$, see Fig.~\ref{fig:unbraiding_closed}. Namely, we find that the nodal structure can be unbraided without closing a neighboring gap, bringing the band structure to a state where nodes of opposite split-biquaternion charges can be annihilated with local perturbations. However, we emphasize that such unbraiding process, involving introduction of additional rings, and refined twisting of the nodal rings [Figs.~\ref{fig:unbraiding_closed}(b)-\ref{fig:unbraiding_closed}(d)] to flip charges on unbraiding enabling further ring annihilation, is highly-non local, effectively providing a protection against arbitrary local perturbations.  We also provide an explicit parametrization of this process in Appendix~\ref{App::C}.

\begin{figure*}[t!]
\centering
    \includegraphics[width=\linewidth]{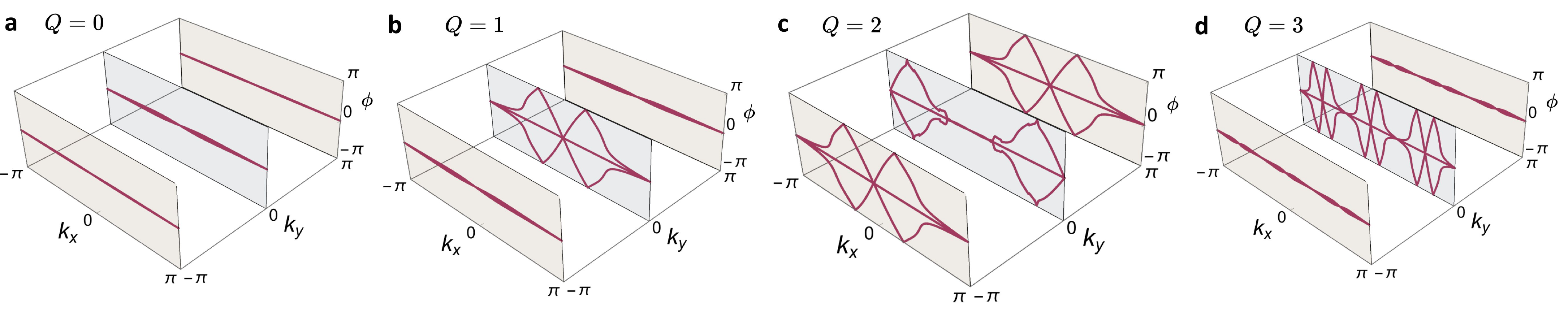}
  \caption{\small Non-Abelian Wilson loop winding for Pontryagin index (\vec{a}) $Q = 0$, (\vec{b}) $Q = 1$, (\vec{c}) $Q=2$, and (\vec{d}) $Q=3$. The loops were evaluated at $k_y = 0$, and high-symmetry planes $k_y = \pi$. For $Q=0$ we find no winding of the Wilson loop eigenvalues $\phi$, along with the winding of $Q=1$, $Q=2$, $Q=3$ being even, contrary to the Stiefel-Whitney insulators with odd Wilson loop winding. We observe that the combined total winding of the eigenvalues within two planes sums to $2Q$.}
  \label{fig:WilsonLoop}
\end{figure*}

Interestingly, we find the nodal rings to enclose two-dimensional regions, which induce $\pi$-phase shifts on parallel-transporting eigenstate frames around them. In other words, such regions are associated with the discontinuities in the Berry connection over the BZ. We refer to these as Dirac sheets, in analogy to the lower-dimensional analogs, namely Dirac strings, corresponding to the gauge connection-discontinuities in two-dimensional Euler phases. In the further parallel-transport study, for each value $Q \neq 0$, we additionally obtain Wilson loop spectrum, showing non-trivial windings across the BZ, contrary to $Q = 0$, see Fig.~\ref{fig:WilsonLoop}. The windings obtained are even, as in Euler phases, contrary to the odd windings found in Stiefel-Whitney insulators~\cite{Ahn2019SW}.

\section{Reference flag limits}\label{sec::V}

In this section, we study the multigap topology of the full flag limit of the Hamiltonian obtained from the models with non-trivial Pontryagin index, when all bands are non-degenerate across the entire BZ. This can be achieved by annihilating the nodes after unbraiding the nodal structure without closing the principal bulk energy gap, as we demonstrated in Fig.~\ref{fig:unbraiding_open}. As all nodes are removed, the classifying space is given by $\mathsf{Fl}_{1,1,1,1}$ and through homotopy classification we obtain the following homotopy classes of Hamiltonians ${\pi_3(\mathsf{Fl}_{1,1,1,1}) \cong \pi_3(SO(4)) \cong \mathbb{Z} \oplus \mathbb{Z}}$. In other words, by removing the degeneracies between bands, a phase transition occurred which allows the new system to host two $\mathbb{Z}$ invariants ($w_L, w_R$) rather than one ($Q$). We find that these invariants are not independent. These flag limits and limits hosting known topologies serve as an important reference to further elucidate our above findings.

As the classifying space has a direct link to $SO(4)$, it is useful to consider how the $\mathbb{Z} \oplus \mathbb{Z}$ invariant arises in the case of $SO(4)$ matrices. In this regard it is useful to call upon the well known isomorphism

\begin{equation}
    SO(4) \cong \frac{S^3_L \times S^3_R}{\mathbb{Z}_2},
\end{equation}
which is a consequence of the fact that $SO(4)$ rotations can be split into two isoclinic rotations (left and right) acting on the vector of interest from the right and left. An arbitrary $SO(4)$ rotation leaves two planes invariant in the sense that any vector within these planes stays in its plane during the rotation. The third homotopy group of this space can then be considered to arise from the 3D winding number of two copies of the three-sphere:

\beq{}
\pi_3(S^3_L \times S^3_R) \cong \pi_3(S^3_L) \oplus \pi_3(S^3_R) \cong \mathbb{Z} \oplus \mathbb{Z}.
\eeq
\\
The $\mathbb{Z}_2$ quotient does not affect $\pi_n ~\text{for}~ n \geq 2$, 
as it is a discrete space. Using this information, a Hamiltonian that hosts this set of invariants may be constructed as detailed in the following.

We start by recognizing that a generic four-band flag Hamiltonian can be factored as
\begin{equation}
    H(\textbf{k}) = V(\textbf{k}) \tilde{E} V(\textbf{k})^{\text{T}},
\end{equation}
where $V$ is an $SO(4)$ matrix of the normalized eigenvectors and $\tilde{E} = \text{diag}[-2,-1,1,2]$. Distinct eigenvalues are used to enforce the fact that there are gaps between all bands. $V$ can then be factored into $V_R V_L$, which are the left and right isoclinic rotation matrices. An explicit form of this factorization reads \cite{perez2017cayley}

\begin{equation}
  V_R=\left(\begin{array}{rrrr}r_0 & -r_3 & r_2 & r_1 \\ r_3 & r_0 & -r_1 & r_2 \\ -r_2 & r_1 & r_0 & r_3 \\ -r_1 & -r_2 & -r_3 & r_0\end{array}\right),  
\end{equation}

\begin{equation}
V_L=\left(\begin{array}{rrrr}l_0 & -l_3 & l_2 & -l_1 \\ l_3 & l_0 & -l_1 & -l_2 \\ -l_2 & l_1 & l_0 & -l_3 \\ l_1 & l_2 & l_3 & l_0\end{array}\right),
\end{equation}
\\
where $r_0^2 + r_1^2 + r_2^2 + r_3^2 = 1$ and $l_0^2 + l_1^2 + l_2^2 + l_3^2 = 1$. These conditions ensure that $V_R$ and $V_L$ are orthogonal matrices and imply that the components $r_i$ and $l_i$ form a pair of four dimensional vectors that lie on three-spheres. 

Considering now the map from the BZ to each of these three-spheres, we can use the winding vector defined in the previous model:

\begin{equation}
  \textbf{r} = (\sin{w_L k_x}, \sin{k_y},\sin{k_z}, 2 - \cos{w_L k_x}-\cos{k_y}-\cos{k_z})^{\text{T}}, 
\end{equation}

\begin{equation}
\textbf{l} = (\sin{w_R k_x}, \sin{k_y},\sin{k_z}, 2 - \cos{w_R k_x}-\cos{k_y}-\cos{k_z})^{\text{T}},  
\end{equation}
\\
to induce the winding on each of these spheres. However, in this case it is possible to have a different winding number on $S^3_R$ and $S^3_L$ through the parameters $w_L$ and $w_R$. The Bloch eigenvectors are then obtained by taking the columns (or rows) of $V_R V_L$.

Although it is not immediately obvious how these two winding numbers can be extracted from the eigenvectors, it is important to note the following. Calculation of the winding number of the Bloch eigenvectors using Eq.~\eqref{Pont_index} gives either $w_L - w_R$ or $w_L + w_R$ depending on if we take $V = V_R V_L$ or $V = V_R V_L^{\text{T}}$. The same number is obtained from all eigenvectors, so this alone is not enough to characterize the phase. 

Further to this, we note that another type of phases described by $\mathcal{PT}$-symmetric four-band models has been proposed recently; namely, the so-called real Hopf insulators \cite{lim2023real}. Contrary to our models, these phases require an occupied and unoccupied two-band subspaces classified in the degenerate limit. In other words, the partitioning in that case corresponds to a classifying space 
\\
\beq{rhi1}
    \widetilde{\mathsf{Gr}}_{2,4}(\mathbb{R}) = SO(4)/[SO(2) \times SO(2)] \cong S^2 \times S^2.
\eeq
\\
Correspondingly, the bulk invariant is given by
\\
\beq{rhi2}
    \pi_3 (\widetilde{\mathsf{Gr}}_{2,4}(\mathbb{R})) \cong \pi_3 (S^2 \times S^2) \cong \mathbb{Z} \oplus \mathbb{Z} \\
\eec
\\
which can be denoted as $(\chi_w, \chi_z)$ and referred to as double Hopf index. The Hopf invariants (indices) are given by \cite{lim2023real}:
\\
\beq{rhi3}
    \chi_{w/z} = -\frac{1}{4\pi} \int_{T^3} a_{w/z} \wedge f_{w/z}
\eec
\\
where $f_{w/z} = \text{d}a_{w/z}$, while $a_{w/z} = i \bar{z}_{w/z} \text{d} \bar{z}_{w/z}$ are connection one-forms defined in terms of Hopf maps induced by the complex vectors $\bar{z}$~\cite{Eulerdrive, lim2023real}.

It follows that the real Hopf insulators with ${(\chi_w, \chi_z) = (0, Q)}$ can be obtained on repartitioning the bands of the model with Pontryagin index, or by closing the upper and lower gaps in the flag limit phases. Any attempts to continuously connect the Hamiltonians of these three phases will necessarily fail, as a gap closing or reopening needs to occur, corresponding to a change of the invariants following from the distinct classifying spaces.

There is, however, an interesting correspondence between $w_L$ and $w_R$ and the invariants constructed to classify the real Hopf insulator in Eqs.~\eqref{rhi1}-\eqref{rhi3}, see Appendix~\ref{App::E},
\\
\begin{equation}
\begin{aligned}
&\chi_w = \frac{1}{2 \pi^2}\int_{\text{BZ}} \dd^3\kv~  \varepsilon_{ijkl}r^i \partial_{k_x} r^j \partial_{k_y} r^k \partial_{k_z} r^l = w_R, \\
&\chi_z = -\frac{1}{2 \pi^2}\int_{\text{BZ}} \dd^3\kv~ \varepsilon_{ijkl}l^i \partial_{k_x} l^j \partial_{k_y} l^k \partial_{k_z} l^l = -w_L. 
\end{aligned}
\end{equation}
\\
We also confirm with further numerical evaluations that $\chi_z$ and $\chi_w$ correspond exactly to the winding number $w_L$ and $w_R$ of the isoclinic rotation matrices. This implies that although these invariants were derived assuming a classifying space of $\widetilde{\mathsf{Gr}}_{2,4}$, the opening of a gap in the top and bottom two-band subspaces does not change the type of topological homotopy invariants. In fact, the only nontrivial change happens on closing two adjacent gaps, as $\pi_3(\mathsf{Fl}_{1,1,1,1}) \cong \pi_3(\mathsf{Fl}_{1,1,2}) \cong \pi_3(\mathsf{Gr}_{2,4}) \cong \mathbb{Z} \oplus \mathbb{Z}$, but $\pi_3(\mathsf{Gr}_{1,4}) \cong \mathbb{Z}$. These results also show that $\chi_w$ and $\chi_z$, as defined above, are not necessarily Hopf invariants, as there is no $S^2$ structure in the classifying space of this model; although there is a link between the winding number on the three-sphere and the Hopf invariant.

We thus retrieve an example of true multigap topology, and a sequence of phase transitions that can be obtained by closing successive gaps. Starting from a fully gapped model with the classifying space $\mathsf{Fl}_{1,1,1,1}$, we can specify $w_L$ and $w_R$ to obtain a ``real Hopf" phase. We can then close two adjacent gaps, leaving only the highest or bottom energy gap open, to obtain the model characterized by the Pontryagin index. Finally, we can trivialize such model by closing the remaining, highest energy gap.

It is important to note that the eigenfunctions in the fully gapped model also possess a 3D winding number (Pontryagin index). The winding, however, can be fully determined from the values of $w_L$ and $w_R$ and therefore does not constitute an independent invariant. The crucial change that happens on closing two adjacent gaps is the ability to mix the bottom three bands through gauge transformations. This introduces a \text{gauge dependence} to $w_L$ and $w_R$, which can be easily checked numerically. It does not, however, affect the winding number of the eigenvectors, as captured by the Pontryagin index, which now becomes an independent invariant. 

We may also consider the stable flag limit in which we extend the band structure to an arbitrary number of isolated bands, enforcing band gaps between them in the flattened Hamiltonian. The classifying space for such a new model is $O(N) / \mathbb{Z}_2^{\times N}$, with third homotopy group, $\pi_3(O(N) / \mathbb{Z}_2^{\times N}) \cong \pi_3(SO(N)) \cong \mathbb{Z}$, for $N \geq 5$. It is known~\cite{fradkin} that every simple compact group $\tilde{G}$ contains an $SU(2)$ subgroup, and the Pontryagin index as defined in equation Eq.~\eqref{eq:Tr3} classifies all the maps $\pi_3(\tilde{G})$. The invariant can actually be evaluated explicitly from Eq.~\eqref{eq:Tr3} by inserting the frame of $N$ eigenvectors in the place of matrix $U$. Alternatively, the $\mathbb{Z}$ invariant arising from the homotopy classification can be understood as the number of times the Hamiltonian wraps the $SU(2)$ subgroup of $SO(N)$.

\section{Edge states and disorder}\label{sec::VI}

In this section we comment on the edge states due to the bulk-boundary correspondence in the reference flag limits, which we contrast with the original models consisting of one principal gap and a three-band subspace. Furthermore, we show that these edge states are robust up to gap-closing disorder in the $3 \oplus 1$ Pontryagin phases, as well as in the related flag limits. 

\subsection{Edge states in flag limits}

\begin{figure}[t!]
\centering
  \includegraphics[width=\columnwidth]{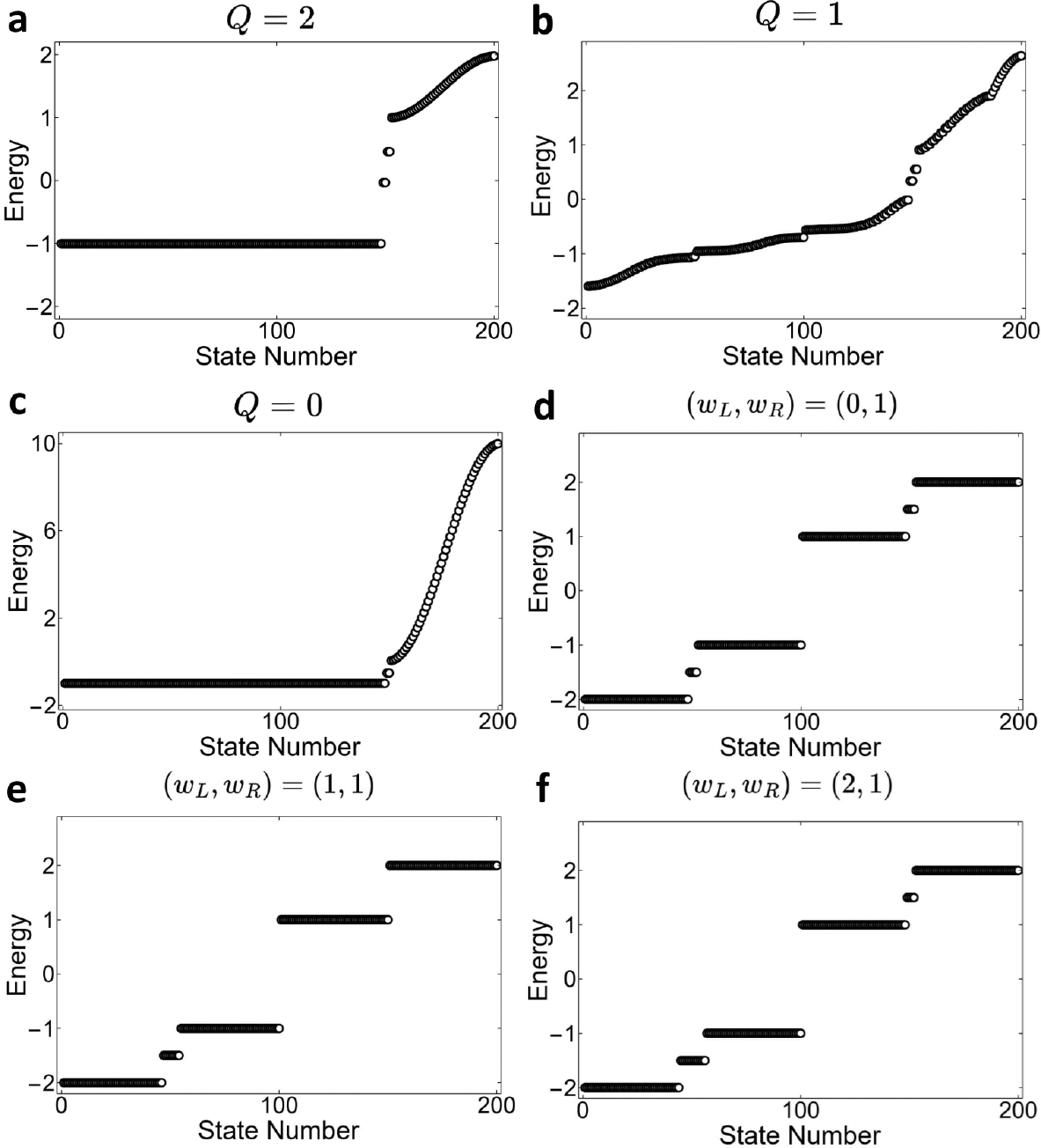}
    \caption{\small Comparison of edge states in the bulk gaps of insulators with Pontryagin indices $Q=2$, $Q=1$, and $Q = 0$, realized in  \textbf{(a)}--\textbf{(c)} the model Hamiltonians~\eqref{Pont3D_H},  (\textbf{d})~flag limit phase $(w_L, w_R) = (
  0,1)$, (\textbf{e}) $(w_L, w_R) = (
  1,1)$, (\textbf{f}) ${(w_L, w_R) = (
  2,1)}$. The edge states are evaluated under open boundary conditions in the $x$ direction with $(k_y, k_z) = (0.5,0.5)$. We find that the bulk-boundary correspondence of these three-dimensional $\mathcal{PT}$-symmetric phases yields the multiplicities of edge states $g = 4Q$ in the principal gap of $3 \oplus 1$ model, and $g_1 = 4(w_L+w_R)$, $g_3 = 4(w_L-w_R)$ in bottom and top gaps in the flag limit. While the topologically trivial $\mathcal{PT}$-symmetric phases with $Q=0$ can host edge states, these are not topologically protected due to the triviality of the bulk invariant, as we verify against disorder (Sec.~\ref{sec::VI}). On closing two lower gaps with a band inversion, $g = g_3$, which can be mapped to the correspondence of invariants. $Q = (w_L-w_R)$. The four-fold multiplicity of edge states can be understood in terms of the interplay of the symmetries: the essential $\mathcal{PT}$, as well as $\mathcal{T}$ ($\mathcal{T}^2 = 1$), and $\mathcal{P}$ admitted by the models.}
\label{fig:edgestates}
\end{figure}

We further elaborate on the bulk-boundary correspondence between the flag invariants $w_L, w_R$ and the presence and degeneracies of the multigap edge states. We find that top and bottom gaps support the presence of dangling edge states, with degeneracies given by multiples of $w_L - w_R$ and $w_L + w_R$, see Fig.~\ref{fig:edgestates}. This should be contrasted with the $Q$ edge states in the principal gap of degenerate $3 \oplus 1$ limit with Pontryagin index, as well as with real Hopf insulators, which do not require a presence of two subsidiary band gaps in occupied and unoccupied two-band subspaces. Additionally, the further connection between edge states in flag limit and $3 \oplus 1$ phases with Pontryagin index can be seen on breaking $\mathcal{T}$, while keeping $\mathcal{PT}$ in the latter, by an addition of a constant matrix term. Namely, this results in the additional edge states appearing in the lower nodal part of the three-band subspace, besides the edge states in the principal gap. As we showed in Sec.~\ref{sec::IV}, the nodes are not protected by the bulk invariant, hence their removal establishes a link of edge states in  $3 \oplus 1$ phase to the edge states in the full flag limit.

\subsection{Robustness to disorder}

In this section, we show the protection of edge states supported by the non-trivial Pontryagin index $Q$ of the bulk Hamiltonian, up to gap closing. While the trivial phase $Q=0$ in the proposed model also has the associated edge modes, we show that these are not topologically protected, hence not robust to disorder. 

\begin{figure}
\centering

\includegraphics[width=\columnwidth]{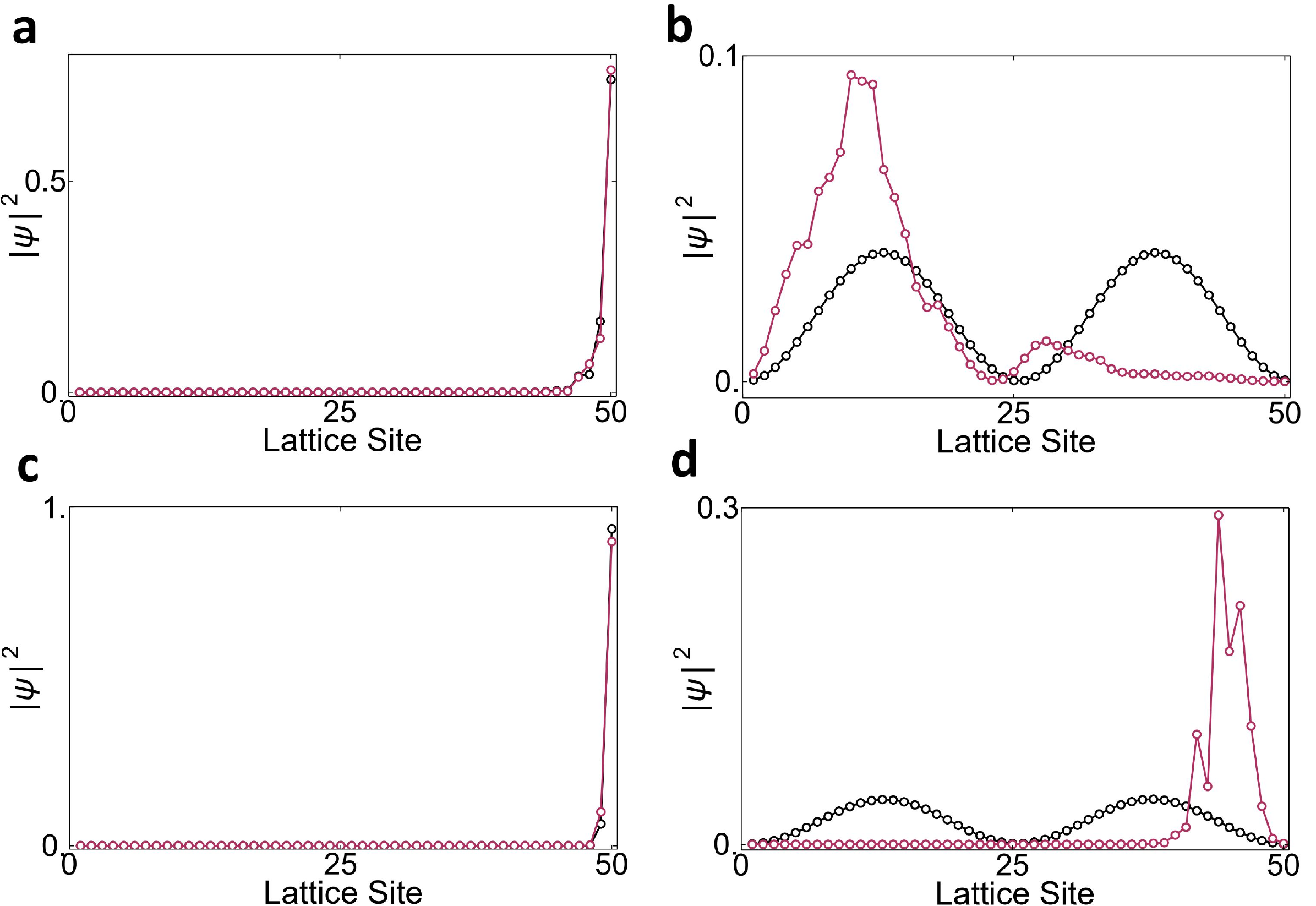}
    \caption{\small Edge and bulk states of (\vec{a})--(\vec{b}) the Pontryagin $Q = 1$ and (\vec{c})--(\vec{d}) associated flag limit $(w_L, w_R) = (0,1)$ phases subject to weak disorder ($W=0.3$). The one-dimensional wavefunction sections of clean phases (\textit{black}) were plotted against the same states perturbed with weak Anderson disorder (\textit{red}). In both phases, the edge states remain exponentially localized  (\vec{a}), (\vec{c}) as long as the disorder strength is not sufficient to close the gap. The bulk states (\vec{b}), (\vec{d}) do not show similar robustness, becoming distorted on adding disorder.}
\label{fig:disorder}
\end{figure}

We impose Anderson disorder by the following perturbation Hamiltonian
\beq{}
    \Delta H_{\text{disorder}} = \sum_{n,i} \delta \mu_{i,n} c^\dagger_{i,n} c_{i,n},
\eeq
where $n$ is a unit cell label, $i = 1,2,3,4$ is an orbital label, and $\delta \mu_{i,n} \in \big[-W, W \big] $ is a local change in the chemical potential on adding disorder with uniform random distribution and amplitude $W$. We find that the edge states remain exponentially localized on adding weak disorder and dissolve at the disorder strength $W$ above the size of the bulk gap, see Fig.~\ref{fig:disorder}. 

\section{Experimental realizations}\label{sec::VII}

We now further elaborate on experimental realizations, which we suggest for studying physical manifestations of the novel type of three-dimensional Pontryagin band topology studied in this work. 

First, we propose a metamaterial realization of the described minimal phases with $Q = 2$, $Q = 1$, $Q = 0$ ($p_x, p_y, p_z = 1$) with the correspondent edge states, hence a way to measure and empirically validate the $\mathbb{Z}$ invariant given by the Pontryagin index. The protocol is based on the idea of extending the known simulations with acoustic resonators~\cite{Jiang_2021,Jiang_meron}, to a 3D synthetic matter construction. To generate the lowest Pontryagin indices, connecting tubes up to the second neighbors is necessary, and we propose that the $\pi/2$ phase shifts generating imaginary hopping amplitudes can be ensured by proper phase-shifting impedance matching in the materials constituting the connecting tubes. On extending the setup of an analogous experiment used to study non-Abelian band topology \cite{Jiang_2021}, the amplitudes of hopping parameters can be controlled with the diameters of the tubes, with coupling of any two connected 
resonators being captured by an effective Hamiltonian  
\\
\beq{}
H_\text{eff} = 
  \begin{pmatrix}
    \om_1 & e^{i\phi} |\kappa|\\
    e^{-i\phi} |\kappa| & \om_2 
    \end{pmatrix}.
\eeq
\\
Here, $|\kappa|$ and $\phi$ correspond to the amplitude and the phase of the coupling $\kappa$ representing particular hopping, and $\om_1, \om_2$ are natural resonator frequencies identified with the onsite energies in the tight-binding model.

While phases with higher index $Q$ can, in principle, be created, that would require an addition of farther-neighbor connections, which might be unfeasible from the technical point of view. We propose that the mass term $m$ crucial for the topological phase transitions can be controlled by changing thickness of the metamaterial tubes. Such procedure would realize a three-dimensional extension of the protocols implemented in the previous works studying Euler topology in two spatial dimensions \cite{Jiang_2021, Jiang_meron}. 

Additionally, we propose that the non-trivial nodal structures can be realized in optical trapped-ion experiments. We would expect the protocol to be analogous to the closely-related experiment used for studying the Euler class in the lower-dimensional, topological Euler insulators \cite{Zhao_2022} with three-band Euler Hamiltonians realized in hyperfine states of ytterbium $^{179}\text{Yb}^{+}$ ions. The four bands with Pontryagin index represented by the fourth-band winding might be realized in atomic states of four-level systems, such as e.g. neodymium $\text{Nd}^{3+}$, in that case with states labeled by term symbols $^4F_{5/2}$, $^4F_{3/2}$, $^4I_{9/2}$, $^4I_{11/2}$. Here, inverting the band structure while not changing topology, such that three bands are unoccupied, might be simpler to achieve in a real experiment. The explicit parametrization of the Hamiltonians naturally realizing the manipulation of the linked nodal structures by braiding processes described in Sec.~\ref{sec::IV}, is provided in Appendix~\ref{App::C}. We note that the unbraiding on closing the principal gap (Fig.~\ref{fig:unbraiding_closed}), which is accessible by a simple diagonal term, might be experimentally simpler to realize than without closing the gap (Fig.~\ref{fig:unbraiding_open}).

\section{Discussion and Conclusion}\label{sec::VIII}

We show that the Pontryagin index naturally induces a $\mathbb{Z}$-type invariant in real-valued three-dimensional four-band Hamiltonians, which we further corroborate by  topological phase transitions and bulk-boundary correspondence. This can be contrasted with other known topologies in three-dimensional systems. Contrary to the stable $\mathbb{Z}_2$ invariant of a three-dimensional topological insulator (Altland-Zirnbauer class AII), protected by  spinful $\mathcal{T}$ ($\mathcal{T}^2 = -1$) symmetry, the centrosymmetric case thereof, or an axion insulator with $\theta = \pi$ and broken $\mathcal{T}$-symmetry; we find that the Pontryagin models host a Chern-Simons three-form for any winding number $Q$, yielding trivial $\theta$, once the gauge freedom in the occupied bands is taken into consideration ($\theta = 4 \pi Q \equiv 0 ~\text{mod}~2\pi$). 

Additionally, we find that the even Wilson loop winding excludes a possibility of existence of 3D Stiefel-Whitney insulators with a $\mathbb{Z}_2$ invariant (i.e.~second Stiefel-Whitney class $w_2$) in the oriented Hamiltonians introduced in this work. By studying the manifestations of the single bulk invariant ($Q \in \mathbb{Z}$), which include the associated bulk-boundary correspondence (Sec.~\ref{sec::IV}), we also deduce that our findings are distinct from the real Hopf insulators hosting surface Chern numbers at the boundaries \cite{lim2023real, alexandradinata2021teleportation}. Hence, by exhaustion of the known three-dimensional $\mathcal{PT}$-symmetric topological phases, we conclude that the Pontryagin-indexed three-band subspace indeed hosts a different type of non-Abelian real topology, which to the best of our knowledge has not been reported in previous works.

Overall, our findings provide a realization of non-Abelian multigap topological insulators with a single $\mathbb{Z}$-invariant in three spatial dimensions, as supported by the multitude of unique results and mathematical relations to other types of topologies presented in this work. We conclude that, contrary to the Abelian topological insulators with a single $\mathbb{Z}$-valued Hopf index, the origin of the invariant stems from the non-triviality of Pontryagin index characterizing the Bloch bundle. We introduced models for arbitrary index, studied the bulk-boundary correspondence due to the integer invariants, investigated the stability of edge modes against disorder, and also referenced these findings to the full flag limit. The index-changing topological phase transitions, ultimately trivializing the model, were studied. Upon addition of the Hamiltonian terms opening all gaps and accessing the full flag limit, we made the connections to real Hopf and axion insulators. The nodal structures, with non-trivial linking numbers removable by highly non-local parallelization of the subbundle corresponding to the three-band subspace hosting the split-biquaternion nodes, were studied, and demonstrated within a class of minimal models. These simple models promise a realization in a wide variety of experimental settings that include metamaterials and quantum simulators. Despite the ``trivializability" of nodal rings discussed in this work, we showed a non-Abelian character of the nodes in a three-dimensional setting, offering non-trivial platform for fusion and braiding, beyond the quaternion algebra of nodes in Euler semimetals.\\

\section{Acknowledgements}

W.~J.~J. acknowledges funding from the Rod Smallwood Studentship at Trinity College, Cambridge. A.~B. has been partly funded by a Marie Sklodowska-Curie fellowship, Grant No. 101025315. R.-J.~S. acknowledges funding from a New Investigator Award, EPSRC Grant No. EP/W00187X/1, as well as Trinity College, Cambridge.

\bibliography{references}

\appendix
\newpage
\section{Pontryagin index in terms of non-Abelian Berry connection}\label{App::A}

Here, we briefly show how the Pontryagin index corresponding to the hyperspherical ($S^3$) winding can be re-expressed in terms of the elements of non-Abelian Berry connection. 

First, we start by noticing that by orthogonality of the normalized eigenvectors constituting vierbeins
\\
\beq{}
    (\vec{u}_4)_i = \varepsilon_{ijkl}(\vec{u}_1)_j(\vec{u}_2)_k(\vec{u}_3)_l
\eec
\\
Equivalently, on defining a coordinate-free notation, where two wedges imply a contraction with $\varepsilon_{ijkl}$, i.e., with the antisymmetric pseudotensor, we have: $\vec{u}_4 = \vec{u}_1 \wedge \vec{u}_2 \wedge \vec{u}_3$. Hence, in such notation, we can rewrite our invariant as
\\
\beq{}
    Q = \frac{1}{2\pi^2} \int_{S^3} \dd^3\kv~ \vec{u}_4 \cdot (\partial_{k_x} \vec{u}_4 \wedge \partial_{k_y} \vec{u}_4 \wedge \partial_{k_z} \vec{u}_4).
\eeq
\\
Using the cyclic property of the scalar product
\\
\beq{}
    Q = \frac{1}{2\pi^2} \int_{S^3} \dd^3\kv~ (\vec{u}_4 \wedge \partial_{k_x} \vec{u}_4 \wedge \partial_{k_y} (\vec{u}_1 \wedge \vec{u}_2 \wedge \vec{u}_3)) \cdot \partial_{k_z} \vec{u}_4.
\eeq
\\
For any five four-vectors, we consistently derive a higher dimensional analog of a triple vector product, within introduced coordinate-free notation
\\
\beq{}
\begin{split}
    \vec{a} \wedge \vec{b} \wedge (\vec{c} \wedge \vec{d} \wedge \vec{e}) =
    \Big[(\vec{a}\cdot\vec{c})(\vec{b}\cdot\vec{d})-(\vec{a}\cdot\vec{d})(\vec{b}\cdot\vec{c})\Big]\vec{e} + \\ + \Big[(\vec{a}\cdot\vec{e})(\vec{b}\cdot\vec{c})-(\vec{a}\cdot\vec{c})(\vec{b}\cdot\vec{e})\Big]\vec{d} + \\ + \Big[(\vec{a}\cdot\vec{d})(\vec{b}\cdot\vec{e})-(\vec{a}\cdot\vec{e})(\vec{b}\cdot\vec{d})\Big]\vec{c},
\end{split}
\eec
\\
which with constraints $\vec{u}_i \cdot \vec{u}_j = \delta_{ij}$ by orthonormality, and $\vec{u}_i \partial_{a} \vec{u}_i = \frac{1}{2} \partial_{a} (\vec{u}_i \cdot \vec{u}_i) $, as well as $\vec{u}_i \partial_{a} \vec{u}_j = - \vec{u}_j \partial_{a} \vec{u}_i$ for any $i = 1,2,3,4$, and $a = k_x,k_y,k_z$ by the reality condition, yields on direct substitution
\\
\beq{}
\begin{split}
    Q = \frac{1}{2\pi^2} \int_{S^3} \dd^3\kv~ \Big[ (\vec{u}_4 \partial_{k_x} \vec{u}_1) (\vec{u}_4 \partial_{k_y} \vec{u}_2) (\vec{u}_4 \partial_{k_y} \vec{u}_3) + \\ - (\vec{u}_4 \partial_{k_x} \vec{u}_1) (\vec{u}_4 \partial_{k_y} \vec{u}_3) (\vec{u}_4 \partial_{k_y} \vec{u}_2) + \dots \Big].
\end{split}
\eeq
\\
Recognizing elements of the non-Abelian Berry connection gives
\\
\beq{}
    Q = \frac{1}{2\pi^2} \int_{T^3} \dd^3\kv~ \Big[ A^x_{41} A^y_{42} A^z_{43} - A^x_{41} A^y_{43} A^z_{42} + \dots \Big].
\eeq
\\
Rewriting in a symmetrized form with respect to the other components (permutations of $k_x$, $k_y$, $k_z$), from which the invariant can also be computed 
\\
\beq{}
    Q = \frac{1}{2\pi^2} \int_{T^3} \dd^3\kv~ \varepsilon_{ijk} A^i_{41} A^j_{42} A^k_{43}
\eec
\\
which completes the proof.

\section{Wannier ladders and surface Wilson loops}\label{App::B}
\begin{figure*}[t!]
     \includegraphics[width=\linewidth]{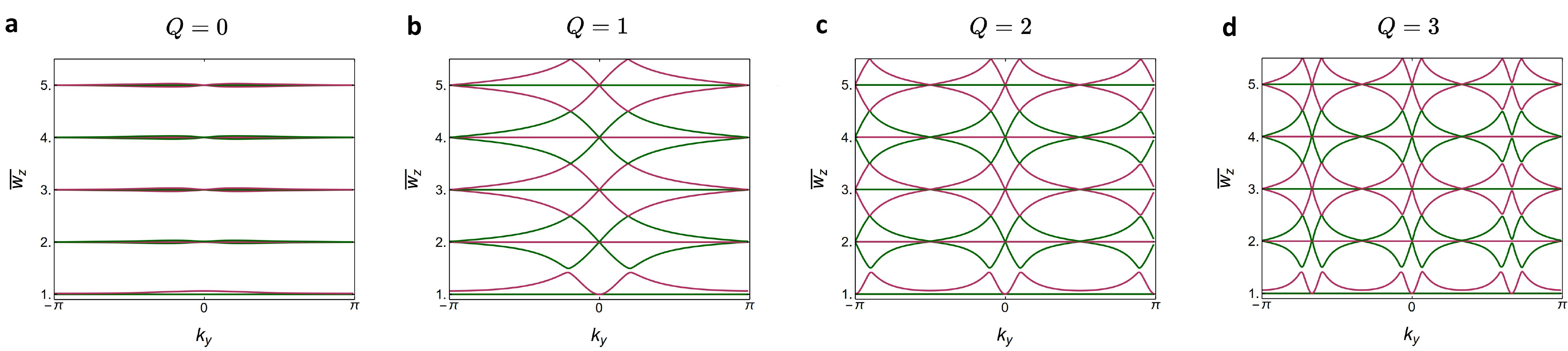}
    \caption{\small Wannier ladders for Pontryagin index (\vec{a}) $Q = 0$, (\vec{b}) $Q = 1$, (\vec{c}) $Q=2$ and (\vec{d}) $Q=3$. The flows of HWF centers ($\bar{w}_z$) were evaluated along $k_y$ on setting $k_x = 0$. For $Q=0$ we find no net flow of HWF centers in the ladder. For non-trivial $Q \in \mathbb{Z}$ (\vec{b})--(\vec{d}), we observe two HWF centers flowing by $\pm 2Q$ real-space unit cells, along with a third HWF center without any flow, all of which correspond to the occupied three-band subspace. On the contrary, the unoccupied band subspace supports no flow in the associated HWF centers.}
    \label{fig:ladders}
\end{figure*}
\begin{figure*}[t!]
     \includegraphics[width=\linewidth]{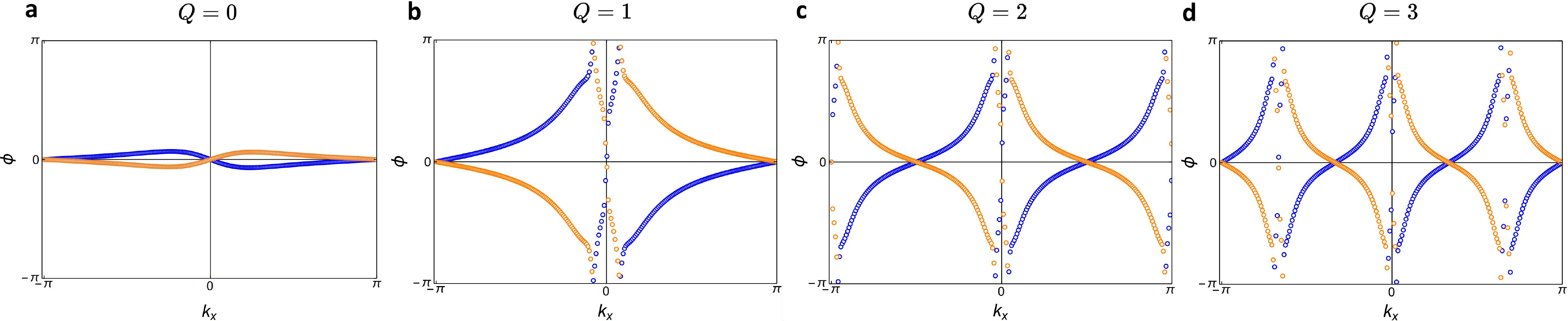}
    \caption{\small Surface Wilson loops for (\vec{a}) $Q = 0$, (\vec{b}) $Q = 1$,  (\vec{c}) $Q=2$ and (\vec{d}) $Q=3$ along $k_x$. Here, the plotted Wilson loop eigenvalues $\phi$ were found for two surface bands associated with two HWF center ($\bar{w}_z$) sheets closest to the surface, on isolating these from the rest of the Wannier ladder. For original HWF sheets, see the Wannier ladder in Fig.~\ref{fig:ladders}. For $Q=0$ we find no Wilson loop winding. For non-trivial $Q \in \mathbb{Z}$ (\vec{b})--(\vec{d}), we find two opposite eigenvalues $\pm 2Q$. While such pair does not carry a surface Chern number, as explained in the main text; under the $\mathcal{C}_2\mathcal{T}$ symmetry satisfied on the surface, such winding is topologically protected against the symmetry-preserving perturbations. Based on the deduced Wilson loop winding, protected under the symmetry enforcing a reality condition~\cite{bouhon2019nonabelian}, we conclude that the surface bands host a surface Euler invariant $\chi_s$, which we find on two opposite surfaces under the open boundary conditions in $z$ direction.}
    \label{fig:SurfaceWilson}
\end{figure*}

Here, we provide further details on the evaluation of the Wannier ladders and the surface Wilson loops used to study the bulk-boundary correspondence of $\mathcal{PT}$-symmetric Hamiltonians with arbitrary Pontryagin index $Q \in \mathbb{Z}$. The Wannier ladders and surface Wilson loops are presented in Fig.~\ref{fig:ladders} and Fig.~\ref{fig:SurfaceWilson}, correspondingly.\\

The hybrid Wannier functions (HWFs), maximally localized in $z$-direction, and used to construct the Wannier ladders, were obtained as the eigenfunctions of the $z$-component of the position operator projected on the occupied subspace. The Hamiltonian is first Fourier transformed in the $z$ direction and placed on a chain with 20 sites under open boundary conditions. The ground state projector is constructed from the occupied energy eigenstates as $\hat{P}=\sum^{\text{occ}}_n \ket{\psi_n}\bra{\psi_n}$. The operator $\hat{P}\hat{z}\hat{P}$ is then numerically diagonalized on a mesh in the 2D reduced Brillouin zone (with momenta $k_x$ and $k_y$) and the eigenvectors and eigenvalues are extracted, corresponding to the HWF and their centers ($\bar{w}_z$) in the $z$-direction respectively.\\

For the surface Wilson loop winding, we employ a standard procedure of computing a non-Abelian Wilson loop, here with surface bands given by $\ket{u^s_{n}(k_x,k_y)}$. As explained in the Fig.~\ref{fig:SurfaceWilson} caption, we use the deduced winding of Wilson loop eigenvalues to exclude the presence of surface Chern numbers ($C_{s,\text{tot}}$), as predicted from an analytical argument. Moreover, we retrieve that the non-trivial winding for bulk $Q\neq 0$ corresponds to the surface Euler topology protected by $\mathcal{C}_2\mathcal{T}$ symmetry. Here, a~pair of surface bands hosts an Euler invariant $\chi_s$ at each surface of the system, contrary to the net opposite surface Chern numbers on the boundaries of real Hopf insulators~\cite{lim2023real}. Finally, we comment that the surface Euler invariant is defined modulo a sign, which can be changed under a gauge transformation $\chi_s \rightarrow -\chi_s$, while the surface Wilson loop winding is preserved. An identical gauge ambiguity is known for the bulk Euler class in two-dimensional materials~\cite{Jiang_2021}.

\section{Parametrizations for unbraiding}\label{App::C}
The trivialization of the nodal structure can be cast in a particularly simple form by adding a perturbation with only on-site energies to the Bloch Hamiltonian. The term used in these calculations is $t~ \text{diag}[-6,0,4,10]$, however, any term of the form $t~ \text{diag}[E_1,E_2,E_3,E_4]$ with the condition $E_1 < E_2 < E_3 < E_4$ will accomplish the desired unbraiding. The parameter $t$ controls the magnitude of the perturbation. In the limit as $t \rightarrow \infty$, the Hamiltonian approaches the form of the perturbation with four flat bands at distinct energies and trivial eigenvectors. This implies that in the process of tuning $t$, the nodal rings must unlink, as this is the only way the occupied bands can develop gaps. The eigenvectors of the limiting Hamiltonian will be constant, as the perturbation is diagonal in the original basis, which leads to all Berry connections (including non-Abelian ones) being identically zero. This implies that any topological invariant constructed from these connections must also be zero, so the bulk gap must close in the process as well to facilitate this phase transition. Both of these effects are observed in the plots shown in Fig. \ref{fig:unbraiding_closed}.

As mentioned in the main text, it is also possible to unbraid the structure without closing the uppermost bulk gap and therefore without trivializing the phase. To accomplish this, we first pick a parallelized basis following Refs.~\cite{Hatcher_1,Hatcher_2} for the (unnormalized) eigenvectors of the occupied subspace as follows:

\begin{widetext}
\begin{align}
\begin{split}
    &\ket{u_1} = (-\sin{k_y},\sin{k_x},-(m-\cos{k_x}-\cos{k_y}-\cos{k_z}),\sin{k_z})^{\text{T}}, \\
    &\ket{u_2} = (-\sin{k_z}, m-\cos{k_x}-\cos{k_y}-\cos{k_z},\sin{k_x},-\sin{k_y})^{\text{T}}, \\
    &\ket{u_3} = (-(m-\cos{k_x}-\cos{k_y}-\cos{k_z}),-\sin{k_z},\sin{k_y},\sin{k_x})^{\text{T}}. \\  
\end{split}
\end{align}  
\end{widetext}

This basis forms a nowhere-vanishing smooth section of $TS^3$. We may then write the Hamiltonian defined in Eq.~\eqref{Pont3D_H} in the spectrally decomposed form with the addition of a diagonal term to reveal the nodal structure:

\begin{equation}
\begin{split}
    H(\kv) = \ket{u_4}&\bra{u_4} - \ket{u_1}\bra{u_1} - \ket{u_2}\bra{u_2} - \\ & -\ket{u_3}\bra{u_3} + ~\text{diag}[-0.6,0,0.4,1].
\end{split}
\end{equation}

Finally, to obtain the unbraiding procedure, we add a tuning parameter $t \in [0,1]$ which disconnects the band associated to $\ket{u_3}$ from the others in the occupied subspace:

\begin{equation}\label{eigendecomp}
\begin{split}
       H(\kv) = \ket{u_4}&\bra{u_4} - \ket{u_1}\bra{u_1} - \ket{u_2}\bra{u_2} - \\ & -(1+t)\ket{u_3}\bra{u_3} + ~\text{diag}[-0.6,0,0.4,1].
\end{split}
\end{equation}
This is equivalent to adding a perturbation to the original Hamiltonian of the form $t \ket{u_3}\bra{u_3}$, with a tuning parameter $t$. It is therefore important to use the unnormalized eigenvectors in this construction, as otherwise the perturbation will not be local in real space. Using these eigenvectors we find that the lack of normalization in Eq.~\eqref{eigendecomp} simply leads to the bands acquiring a dispersion and no change in topology occurs. 

\section{Split-biquaternion nodal algebra}\label{App::D}

In this Appendix, we detail the split-biquaternion charges that can be assigned to the collections of nodes acting differently on parallel-transported vierbeins. The nodes correspond to the band crossings introduced in the multigap flag limit with four isolated bands. We note that the classification is algebraically reminiscent to the one introduced in the previous work \cite{bouhonGeometric2020}.
\\
\begin{figure}[H]
     \includegraphics[width=\columnwidth]{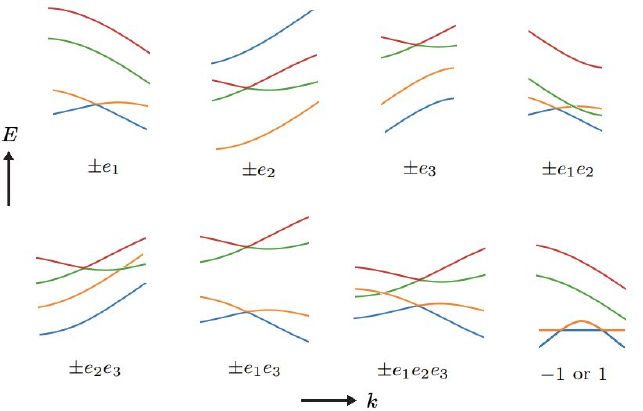}
    \caption{Assigning the non-Abelian charges to different node configurations. The generators of the group are assigned to single nodes in each of the gaps and more complex configurations are achieved by multiplying the charges of the constituent nodes. Note that $e_1 e_2 e_3$ and $-e_1 e_2 e_3$ are in different conjugacy classes, as there is no braiding procedure that can flip the sign of only one of the nodes.}
    \label{fig:nodes}
\end{figure}

\section{Equivalence of $\widetilde{\mathsf{Fl}}_{1,1,1,1}$ winding numbers and double real Hopf invariant}\label{App::E}

In the following, we show the equivalence of the winding numbers generating topology in real Hamiltonians classified by $\widetilde{\mathsf{Fl}}_{1,1,1,1}$ and real Hopf invariants corresponding to winding induced by mapping $S^3 \times S^3 \rightarrow \widetilde{\mathsf{Gr}}_{2,4} \cong S^2 \times S^2$. Throughout this section we denote partial derivatives $\frac{\partial f}{\partial x}$ as $f_x$. The explicit form of the matrix of Bloch eigenvectors in terms of the components of $\textbf{r} = (a,b,c,d)^{\text{T}}$ and $\textbf{l} = (p,q,r,s)^{\text{T}}$ is:
\\
\begin{widetext}
\beq{}
\begin{split}
\left(
\begin{array}{cccc}
 a p+b q-c r-d s & -a s+b r+c q-d p & a r+b s+c p+d q & a q-b p+c s-d r \\
 a s+b r+c q+d p & a p-b q+c r-d s & -a q-b p+c s+d r & a r-b s-c p+d q \\
 -a r+b s-c p+d q & a q+b p+c s+d r & a p-b q-c r+d s & a s+b r-c q-d p \\
 -a q+b p+c s-d r & -a r-b s+c p+d q & -a s+b r-c q+d p & a p+b q+c r+d s \\
\end{array}
\right).
\end{split}
\eeq
\end{widetext}

Taking the rows to be the eigenvectors sorted top to bottom as $(u_1, u_2, u_3, u_4)$, it is possible to compute the Euler connection and Euler curvature using these components. The results are the connections for the valence subspace and the conduction subspace:\\

\begin{widetext}
\beq{}
\begin{split}
a^c_x = (a   p  +b   q  -c   r  +d   s  ) (s   a_{k_x}  +a   s_{k_x}  -r   b_{k_x}  -b   r_{k_x}  -q   c_{k_x}  -c   q_{k_x}  -p   d_{k_x}  -d   p_{k_x})   \\ +(-a   s  -b   r  -c   q  +d   p  ) (p   a_{k_x}  +a   p_{k_x}  -q   b_{k_x}  -b   q_{k_x}   +r   c_{k_x}  +c   r_{k_x} +s   d_{k_x}  +d   s_{k_x}  ) \\ +(a   q  -b   p  -c   s  -d   r  ) (-r   a_{k_x}  -a   r_{k_x}  -s   b_{k_x}  -b   s_{k_x}  +p   c_{k_x}  +c   p_{k_x}  -q   d_{k_x}  -d   q_{k_x}  ) \\ +(-a   r  +b   s  -c   p  -d   q  ) (-q   a_{k_x}  -a   q_{k_x}  -p   b_{k_x}  -b   p_{k_x}  -s   c_{k_x}  -c   s_{k_x}  +r   d_{k_x}  +d   r_{k_x}  ),\\
\end{split}
\eeq
\end{widetext}

\begin{widetext}
\beq{}
\begin{split}
    a^v_x = (a    p   -b    q   -c    r   -d    s   )(-s    a_{k_x}   -a    s_{k_x}   -r    b_{k_x}   -b    r_{k_x}   +q    c_{k_x}   +c    q_{k_x}   -p    d_{k_x}   -d    p_{k_x}) \\ +(a    q   +b    p   -c    s   +d    r   ) (r    a_{k_x}   +a    r_{k_x}   -s    b_{k_x}   -b    s_{k_x}   -p    c_{k_x}   -c    p_{k_x}   -q    d_{k_x}   -d    q_{k_x}   ) \\+(a    r   +b    s   +c    p   -d    q   ) (-q    a_{k_x}   -a    q_{k_x}  +p    b_{k_x}   +b    p_{k_x}   -s    c_{k_x}   -c    s_{k_x}   -r    d_{k_x}   -d    r_{k_x}   ) \\+(a    s   -b    r   +c    q   +d    p   ) (p    a_{k_x}   +a    p_{k_x}   +q    b_{k_x}   +b    q_{k_x}   +r    c_{k_x}   +c    r_{k_x}   -s    d_{k_x}   -d    s_{k_x}  ),
\end{split}
\eeq
\end{widetext}

and similarly for $k_y$ and $k_z$ on simple replacement of $k_x$ in the above expression.\\

The corresponding curvatures of top two (conduction) and bottom (valence) two bands are:\\

\begin{widetext}
\beq{}
\begin{split}
F^c_{xy} =\left(p   a_{k_x}  +q   b_{k_x}  -r   c_{k_x}  +s   d_{k_x}  +a   p_{k_x}  +b   q_{k_x}  -c   r_{k_x}  +d   s_{k_x}  \right)  \left(s   a_{k_y}  -r   b_{k_y}  -q   c_{k_y}  -p   d_{k_y}  -d   p_{k_y}  -c   q_{k_y}  -b   r_{k_y}  +a   s_{k_y}  \right)- \\ \left(p   a_{k_y}  +q   b_{k_y}  -r   c_{k_y}  +s   d_{k_y}  +a   p_{k_y}  +b   q_{k_y}  -c   r_{k_y}  +d   s_{k_y}  \right)  \left(s   a_{k_x}  -r   b_{k_x}  -q   c_{k_x}  -p   d_{k_x}  -d   p_{k_x}  -c   q_{k_x}  -b   r_{k_x}  +a   s_{k_x}  \right)-\\ \left(-s   a_{k_y}  -r   b_{k_y}  -q   c_{k_y}  +p   d_{k_y}  +d   p_{k_y}  -c   q_{k_y}  -b   r_{k_y}  -a   s_{k_y}  \right)  \left(p   a_{k_x}  -q   b_{k_x}  +r   c_{k_x}  +s   d_{k_x}  +a   p_{k_x}  -b   q_{k_x}  +c   r_{k_x}  +d   s_{k_x}  \right)-\\ \left(q   a_{k_y}  -p   b_{k_y}  -s   c_{k_y}  -r   d_{k_y}  -b   p_{k_y}  +a   q_{k_y}  -d   r_{k_y}  -c   s_{k_y}  \right)  \left(-r   a_{k_x}  -s   b_{k_x}  +p   c_{k_x}  -q   d_{k_x}  +c   p_{k_x}  -d   q_{k_x}  -a   r_{k_x}  -b   s_{k_x}  \right)-\\ \left(-r   a_{k_y}  +s   b_{k_y}  -p   c_{k_y}  -q   d_{k_y}  -c   p_{k_y}  -d   q_{k_y}  -a   r_{k_y}  +b   s_{k_y}  \right)  \left(-q   a_{k_x}  -p   b_{k_x}  -s   c_{k_x}  +r   d_{k_x}  -b   p_{k_x}  -a   q_{k_x}  +d   r_{k_x}  -c   s_{k_x}  \right)+\\ \left(-s   a_{k_x}  -r   b_{k_x}  -q   c_{k_x}  +p   d_{k_x}  +d   p_{k_x}  -c   q_{k_x}  -b   r_{k_x}  -a   s_{k_x}  \right)  \left(p   a_{k_y}  -q   b_{k_y}  +r   c_{k_y}  +s   d_{k_y}  +a   p_{k_y}  -b   q_{k_y}  +c   r_{k_y}  +d   s_{k_y}  \right)+\\ \left(q   a_{k_x}  -p   b_{k_x}  -s   c_{k_x}  -r   d_{k_x}  -b   p_{k_x}  +a   q_{k_x}  -d   r_{k_x}  -c   s_{k_x}  \right)  \left(-r   a_{k_y}  -s   b_{k_y}  +p   c_{k_y}  -q   d_{k_y}  +c   p_{k_y}  -d   q_{k_y}  -a   r_{k_y}  -b   s_{k_y}  \right)+\\ \left(-r   a_{k_x}  +s   b_{k_x}  -p   c_{k_x}  -q   d_{k_x}  -c   p_{k_x}  -d   q_{k_x}  -a   r_{k_x}  +b   s_{k_x}  \right)  \left(-q   a_{k_y}  -p   b_{k_y}  -s   c_{k_y}  +r   d_{k_y}  -b   p_{k_y}  -a   q_{k_y}  +d   r_{k_y}  -c   s_{k_y}  \right),
\end{split}
\eeq
\end{widetext}

\begin{widetext}
\beq{}
\begin{split}
F^v_{xy} = - \left(p  a_{k_x} +q  b_{k_x} +r  c_{k_x} -s  d_{k_x} +a  p_{k_x} +b q_{k_x} +c  r_{k_x} -d  s_{k_x}  \right)   \left(s  a_{k_y} -r  b_{k_y} +q  c_{k_y} +p  d_{k_y} +d  p_{k_y} +c  q_{k_y} -b  r_{k_y} +a  s_{k_y} \right)- \\ \left(p  a_{k_y} -q  b_{k_y} -r  c_{k_y} -s  d_{k_y} +a  p_{k_y} -b  q_{k_y} -c  r_{k_y} -d  s_{k_y} \right)  \left(-s  a_{k_x} -r  b_{k_x} +q  c_{k_x} -p  d_{k_x} -d  p_{k_x} +c  q_{k_x} -b  r_{k_x} -a  s_{k_x} \right)- \\ \left(q  a_{k_y} +p  b_{k_y} -s  c_{k_y} +r  d_{k_y} +b  p_{k_y} +a  q_{k_y} +d  r_{k_y} -c  s_{k_y} \right)  \left(r  a_{k_x} -s  b_{k_x} -p  c_{k_x} -q  d_{k_x} -c  p_{k_x} -d  q_{k_x} +a  r_{k_x} -b  s_{k_x} \right)-\\ \left(r  a_{k_y} +s  b_{k_y} +p  c_{k_y} -q  d_{k_y} +c  p_{k_y} -d  q_{k_y} +a  r_{k_y} +b  s_{k_y} \right)  \left(-q  a_{k_x} +p  b_{k_x} -s  c_{k_x} -r  d_{k_x} +b  p_{k_x} -a  q_{k_x} -d  r_{k_x} -c  s_{k_x} \right)+ \\ \left(p  a_{k_x} -q  b_{k_x} -r  c_{k_x} -s  d_{k_x} +a  p_{k_x} -b  q_{k_x} -c  r_{k_x} -d  s_{k_x} \right)  \left(-s  a_{k_y} -r  b_{k_y} +q  c_{k_y} -p  d_{k_y} -d  p_{k_y} +c  q_{k_y} -b  r_{k_y} -a  s_{k_y} \right)+ \\ \left(q  a_{k_x} +p  b_{k_x} -s  c_{k_x} +r  d_{k_x} +b  p_{k_x} +a  q_{k_x} +d  r_{k_x} -c  s_{k_x} \right)  \left(r  a_{k_y} -s  b_{k_y} -p  c_{k_y} -q  d_{k_y} -c  p_{k_y} -d  q_{k_y} +a  r_{k_y} -b  s_{k_y} \right)+ \\ \left(r  a_{k_x} +s  b_{k_x} +p  c_{k_x} -q  d_{k_x} +c  p_{k_x} -d  q_{k_x} +a  r_{k_x} +b  s_{k_x} \right)  \left(-q  a_{k_y} +p  b_{k_y} -s  c_{k_y} -r  d_{k_y} +b  p_{k_y} -a  q_{k_y} -d  r_{k_y} -c  s_{k_y} \right)+ \\ \left(s  a_{k_x} -r  b_{k_x} +q  c_{k_x} +p  d_{k_x} +d  p_{k_x} +c  q_{k_x} -b  r_{k_x} +a  s_{k_x} \right)   \left(p  a_{k_y} +q  b_{k_y} +r  c_{k_y} -s  d_{k_y} +a  p_{k_y} +b  q_{k_y} +c  r_{k_y} -d  s_{k_y} \right).
\end{split}
\eeq
\end{widetext}

The Hopf invariants can be brought to the form \cite{lim2023real}:

\begin{equation}
\begin{aligned} 
& -\frac{1}{16 \pi^2} \int_{\text{BZ}} \mathrm{a}^c \wedge \mathrm{Eu}^c+\mathrm{a}^v \wedge \mathrm{Eu}^v + \mathrm{a}^c \wedge \mathrm{Eu}^v+\mathrm{a}^v \wedge \mathrm{Eu}^c = \chi_z, \\ 
& -\frac{1}{16 \pi^2} \int_{\text{BZ}}  \mathrm{a}^c \wedge \mathrm{Eu}^c+\mathrm{a}^v \wedge \mathrm{Eu}^v -\mathrm{a}^c \wedge \mathrm{Eu}^v - \mathrm{a}^v \wedge \mathrm{Eu}^c = \chi_w .
\end{aligned}
\end{equation}

Plugging in these expressions and greatly simplifying leads to the following:

\begin{widetext}
\beq{}
\begin{split}
\chi_z = -\frac{1}{16 \pi^2}\int_{\text{BZ}}~\dd^3\kv~  (-8 d  b_{k_z}  a_{k_y}  c_{k_x} +8 d  a_{k_z}  b_{k_y}  c_{k_x} -8 a  d_{k_z}  b_{k_y}  c_{k_x} -8 b  a_{k_z}  d_{k_y}  c_{k_x} \\+ 8 a  b_{k_z}  d_{k_y}  c_{k_x} +8 b  a_{k_y}  d_{k_z}  c_{k_x} -8 b  d_{k_z}  c_{k_y}  a_{k_x} -8 c  b_{k_z}  d_{k_y}  a_{k_x} -\\8 c  d_{k_z}  a_{k_y}  b_{k_x} -8 d  a_{k_z}  c_{k_y}  b_{k_x} -8 a  c_{k_z}  d_{k_y}  b_{k_x} +8 d  a_{k_x}  b_{k_z}  c_{k_y} +8 d  a_{k_y}  b_{k_x}  c_{k_z} -8 d  a_{k_x}  b_{k_y}  c_{k_z} -8 b  c_{k_z}  a_{k_y}  d_{k_x} -8 c  a_{k_z}  b_{k_y}  d_{k_x} +\\8 c  a_{k_y}  b_{k_z}  d_{k_x} +8 b  a_{k_z}  c_{k_y}  d_{k_x} -8 a  b_{k_z}  c_{k_y}  d_{k_x} +8 a  b_{k_y}  c_{k_z}  d_{k_x} +8 c  a_{k_z}  b_{k_x}  d_{k_y} +8 b  a_{k_x}  c_{k_z}  d_{k_y} +8 c  a_{k_x}  b_{k_y}  d_{k_z} +8 a  b_{k_x}  c_{k_y}  d_{k_z}), 
\end{split}
\eeq
\end{widetext}

\begin{widetext}
\beq{}
\begin{split}
    \chi_w = -\frac{1}{16 \pi^2}\int_{\text{BZ}} \dd^3\kv~ (-8 q  s_{k_z}  p_{k_y}  r_{k_x} -8 s  p_{k_z}  q_{k_y}  r_{k_x} +8 s  p_{k_y}  q_{k_z}  r_{k_x} +8 q  p_{k_z}  s_{k_y}  r_{k_x} \\  -8 p  q_{k_z}  s_{k_y}  r_{k_x} +8 p  q_{k_y}  s_{k_z}  r_{k_x} -8 r  s_{k_z}  q_{k_y}  p_{k_x} -8 s  q_{k_z}  r_{k_y}  p_{k_x} \\ -8 q  r_{k_z}  s_{k_y}  p_{k_x} -8 s  r_{k_z}  p_{k_y}  q_{k_x} -8 p  s_{k_z}  r_{k_y}  q_{k_x} -8 r  p_{k_z}  s_{k_y}  q_{k_x} +8 s  p_{k_z}  q_{k_x}  r_{k_y} +8 s  p_{k_x}  q_{k_y}  r_{k_z} -8 r  q_{k_z}  p_{k_y}  s_{k_x} +8 r  p_{k_z}  q_{k_y}  s_{k_x} \\ -8 p  r_{k_z}  q_{k_y}  s_{k_x} -8 q  p_{k_z}  r_{k_y}  s_{k_x} +8 p  q_{k_z}  r_{k_y}  s_{k_x} +8 q  p_{k_y}  r_{k_z}  s_{k_x}  +8 r  p_{k_x}  q_{k_z}  s_{k_y} +8 p  q_{k_x}  r_{k_z}  s_{k_y} +8 r  p_{k_y}  q_{k_x}  s_{k_z} +8 q  p_{k_x}  r_{k_y}  s_{k_z}). 
\end{split}
\eeq
\end{widetext}

Written as components of \textbf{r} and \textbf{l}, the above equations reduce to:
\begin{equation}
\begin{aligned}
&\chi_w = \frac{1}{2 \pi^2}\int_{\text{BZ}} \dd^3\kv~  \varepsilon_{ijkl}r^i r_{k_x}^j r_{k_y}^k r_{k_z}^l, \\
&\chi_z = -\frac{1}{2 \pi^2}\int_{\text{BZ}} \dd^3\kv~ \varepsilon_{ijkl}l^i l_{k_x}^j l_{k_y}^k l_{k_z}^l, 
\end{aligned}
\end{equation}
which are (up to a sign), the 3D winding numbers of the vectors \textbf{r} and \textbf{l}. These winding numbers were defined to be $w_L$ and $w_R$, which completes the proof of correspondence to real Hopf invariants.

\end{document}